\documentclass[12pt ]{iopart}
\expandafter\let\csname equation*\endcsname\relax
\expandafter\let\csname endequation*\endcsname\relax
\usepackage{amsmath}
\usepackage{amsbsy,amssymb}
\usepackage{appendix}
\usepackage{graphicx}
\usepackage{float}
\usepackage{subfig}

\begin{document}

\title[Two-fluid effects on RFP sawtooth relaxation process]{Two-fluid effects on the nonlinear dynamics of RFP relaxation}
\author{Wentan Yan$^{1}$, Ping Zhu$^{2,3}$, Hong Li$^{1}$, Wandong Liu$^{1}$, Bing Luo$^{4}$}
\address{$^1$ KTX Laboratory and Department of Plasma Physics and Fusion Engineering, University of Science and Technology of China, Hefei, Anhui 230026, China}
\address{$^2$  State Key Laboratory of Advanced Electromagnetic Engineering and Technology, International Joint Research Laboratory of Magnetic Confinement Fusion and Plasma Physics, School of Electrical and Electronic Engineering, Huazhong University of Science and Technology, Wuhan, Hubei 430074, China}
\address{$^3$ Department of Nuclear Engineering and Engineering Physics,University of Wisconsin-Madison, Madison, Wisconsin 53706, USA}
\address{$^4$ Hunan Province Key Laboratory Integration and Optical Manufacturing Technology, College of Mathematics and Physics, Hunan University of Arts and Science, Changde, Hunan 415000, China}
\ead{zhup@hust.edu.cn, honglee@ustc.edu.cn}
\vspace{10pt}
\begin{indented}
\item[] Date: 20 August 2026
\end{indented}

\begin{abstract}
	This study  investigates the role of two-fluid effects during magnetic relaxation in reversed-field pinch (RFP) plasmas. Within the multiple-helicity (MH) regime, two-fluid simulations produce distinct sawtooth oscillations, in contrast to the  sawtooth-free state obtained in single-fluid  simulations. Analysis of the magnetic field aligned projection of Faraday's law reveals that, tearing modes collectively generate a dynamo electric field that sustains the magnetic relaxation, a process analogous to the flux-pumping in tokamaks.  Despite the reduced linear tearing-mode growth rates, stronger two-fluid effects produce more pronounced sawtooth activity over the parameter range considered. Modal energy analysis shows that Hall-mediated nonlinear energy redistribution disrupts the coherent tearing-mode dynamics required to sustain steady flux-pumping, thereby facilitating intermittent reconnection. This transition is interpreted as a Hall-mediated dynamical bifurcation between steady flux-pumping and quasi-periodic sawtooth relaxation.


\end{abstract}

\maketitle

\section{Introduction}

Experiments on reversed-field pinch (RFP) plasmas have suggested the non-negligible role of two-fluid effects during relaxation events \cite{Ding-2006,Kuritsyn-2009,King-2012}. For example, Direct measurements in MST, obtained  by correlating fluctuations in current density and magnetic field, show the Hall Dynamo is small at the far edge but becomes significant enough to balance Ohm's law near the magnetic field reversal surface \cite{Shen-1993}. Further core measurements using Faraday rotation diagnostics reveal that the Hall dynamo is spatially localized around resonant surfaces and amplifies dramatically during sawtooth crashes \cite{Ding-2004,Ding-2006}.  Experiments on the TPE-1RM20 device further  demonstrate that the diamagnetic dynamo dominates in collisional regimes \cite{Ji-1995-eff}, indicating that diamagnetic drift  governs the correlation between electron flow and the magnetic field. Observations of rapid momentum transport, evidenced by the flattening of toroidal and poloidal rotation profiles during relaxation events \cite{Kuritsyn-2009}, are consistent with a two-fluid extension of Taylor’s relaxation theory, which predicts that the parallel plasma momentum evolves toward a relaxed state \cite{Steinhauer-1997}, a result that has also been confirmed in extended MHD simulations \cite{King-2012,Sauppe-2017}.

Despite substantial experimental and numerical evidence demonstrating the importance of two-fluid effects, their role in determining the sawtooth onset threshold in RFP plasmas remains less understood. Tokamak studies have shown that two-fluid physics is not only crucial for reproducing the rapid crash time of experimentally observed sawteeth \cite{Yu-2015}, but also fundamentally modifies the onset and nonlinear evolution of the instability \cite{Federico-2011,Yu-2024}. In contrast, corresponding studies in RFP plasmas are  relatively limited. In particular, it is  unclear how two-fluid effects influence  the collective dynamo process that governs the magnetic self-organization  and relaxation in RFPs, or whether these effects favor the maintenance of a quiescent sawtooth-free state or instead driving the system toward repetitive sawtooth oscillations. The extension of resistive MHD model to include the two-fluid effects is therefore essential for predicting the onset threshold  of the nonlinear relaxation dynamics in RFP plasmas.

In this work, we perform systematic three-dimensional nonlinear simulations using the NIMROD code to compare single-fluid and two-fluid RFP behaviors, focusing on sawtooth dynamics  and self-organization. We examine on how two-fluid physics may significantly change the onset threshold of sawtooth activity. For the same dissipation parameters, we find that single-fluid simulations demonstrate the relaxation process toward a shallow-reversal, sawtooth-free multi-helicity state sustained by a quasi-linear MHD dynamo, whereas two-fluid simulations exhibit quasi-periodic sawtooth oscillations. By varying the strength of the two-fluid effects and comparing with the linear tearing mode growth rate, we show that this behavior is not determined by the linear instabilities. Instead, we identify strong Hall-mediated nonlinear interactions as the dominant process governing the sawtooth behavior.


The remainder of this paper is organized as follows. Section 2 introduces the two-fluid MHD model implemented in NIMROD and outlines the simulation setup. Section 3 compares the single-fluid and two-fluid sawtooth dynamics, examines their dependence on the ion-mass scaling factor $m_i$, and tests their sensitivity to the thermal-conductivity model. Section 4 analyzes flux-pumping in the MH state and compares it with its tokamak counterpart. Section 5 examines the radial profiles and helical structures of the MHD and Hall dynamo fields. Section 6 uses a modal-energy decomposition to show how Hall-mediated nonlinear mode coupling redistributes energy among modes, disrupts flux-pumping, and drives quasi-periodic sawtooth relaxation. Finally, section 7 presents the conclusions and discussion.
 
 
\section{Simulation Model and Setup}
We use the three-dimensional nonlinear  two-fluid MHD model implemented in the  NIMROD code  \cite{Sovinec-2003,Glasser-1999}  to simulate the sawtooth process in an RFP plasma.  The  two-fluid  equations are
\begin{eqnarray}
\frac{\partial N}{\partial t} &=& -\nabla\cdot(N\boldsymbol{v}) + \nabla\cdot(D_N\nabla N) \\
\rho\frac{\partial\boldsymbol{v}}{\partial t} &=& -\rho\boldsymbol{v}\cdot\nabla\boldsymbol{v} + \boldsymbol{J}\times\boldsymbol{B} - \nabla P- \nabla \cdot\Pi_{gyro} + \rho\nu\nabla^2\boldsymbol{v} \\
\frac{\partial\boldsymbol{B}}{\partial t} &=& -\nabla\times \boldsymbol{E}  \\
\frac{\partial T}{\partial t} &=& -\boldsymbol{v}\cdot \nabla T - (\gamma-1)T\nabla\cdot \boldsymbol{v} + \nabla\cdot(\boldsymbol{\chi}\cdot \nabla T) \\
\boldsymbol{E} &=& -\boldsymbol{v}\times\boldsymbol{B}+ \eta \boldsymbol{J} +\frac{\boldsymbol{J}\times\boldsymbol{B}}{ne} -\frac{\nabla p_e}{ne}+\frac{m_e}{ne^2}\frac{\partial \boldsymbol{J}}{\partial t}
\end{eqnarray}while $N$, $\rho$,$\boldsymbol{v}$, $P$, $T$,  $\boldsymbol{J}$, $\boldsymbol{B}$, and $\boldsymbol{E}$ represent plasma  number density,  mass density, velocity, pressure, temperature,  current  density, magnetic field and electric field, respectively. $D_n$, $\nu$ and $\boldsymbol{\chi}$ are the coefficients for density  diffusivity, viscosity, and anisotropic thermal conductivity tensor, respectively.  $\eta$ is the resistivity and $\gamma$ is the adiabatic index. The  resistive and viscous dissipations is characterized by  dimensionless Lundquist number $S=\tau_R/\tau_A$ and Reynolds number $M=\tau_\nu/\tau_A$, where $\tau_A=a\sqrt{\mu_0 m_in}/B$ represents the Alfvén time scale, $\tau_R=\mu_0 a^2/\eta_0$ represents the resistive diffusion time scale, and $\tau_\nu=a^2/\nu$ represents the viscous dissipation time.  All computations presented here have a Lundquist number, $S=1.2\times 10^4$, which is roughly 2 orders of magnitude smaller than in the experiment, as limited by accessible computational  resources. 

The two-fluid physics in the model are represented by the Hall and diamagnetic terms in Ohm's law and the gyroviscous stress arising from the finite Larmor radius effects. The gyroviscous stress tensor, derived for a collisional plasma in the small ion gyroradius limit \cite{Ramos-2005}, is given by
\begin{eqnarray}
\boldsymbol{\Pi}_{\text{gyro}} = \frac{m_i p_i}{4 e B} \left[ \boldsymbol{e}_b \times \boldsymbol{W} \cdot \left( \boldsymbol{I} + 3 \boldsymbol{e}_b \boldsymbol{e}_b \right) - \left( \boldsymbol{I} + 3 \boldsymbol{e}_b \boldsymbol{e}_b \right) \cdot \boldsymbol{W} \times \boldsymbol{e}_b \right],
\end{eqnarray}
where $\boldsymbol{W} = \nabla \boldsymbol{v} + (\nabla \boldsymbol{v})^T - \frac{2}{3} \boldsymbol{I} \nabla \cdot \boldsymbol{v}$ is the rate of strain tensor, and $\boldsymbol{e}_b$ is the unit vector along the magnetic field. The strength of the FLR viscous effect is proportional to the ion mass $m_i$, as are the Hall and diamagnetic terms. After substituting the momentum equation without the viscous term, the generalized Ohm's law can be expressed as \cite{Ahedo-2009}
\begin{equation}
\boldsymbol{E} = -\boldsymbol{v} \times \boldsymbol{B} + \eta \boldsymbol{J} + \frac{1}{en} \left( \boldsymbol{J} \times \boldsymbol{B} - \nabla p_e \right) \approx -\boldsymbol{v} \times \boldsymbol{B} + \eta \boldsymbol{J} + \frac{m_i}{e} \left( \frac{D \boldsymbol{v}}{Dt} + \frac{\nabla p_i}{\rho} \right)
\end{equation}
In summary, both categories of two-fluid effects are proportional to $m_i$. By adjusting the magnitude of $m_i$, we can control the strength of different two-fluid effects in our simulations. To avoid altering the plasma density and Alfvén time by a real change in $m_i$, we introduce a scaling factor $m_i/m_{i0}$ in front of the FLR and Hall terms in the code to control the strength of two-fluid effects. In addition to the single-fluid case, we have prepared  three groups of two-fluid cases with various ion mass scaling factors of $1/4$, $1$, and $4$. The  $\beta=0.03$ and electron temperature are kept constant. The corresponding ion skin depth, defined as $d_i = \sqrt{{m_i}/{\mu_0 n e^2}}$, varies from $0.135a$ to $0.54a$.  Additionally, the electron mass is artificially increased by a factor of 10 for numerical convenience.

\begin{figure}[htbp]
\centering
\includegraphics[width=0.6\textwidth]{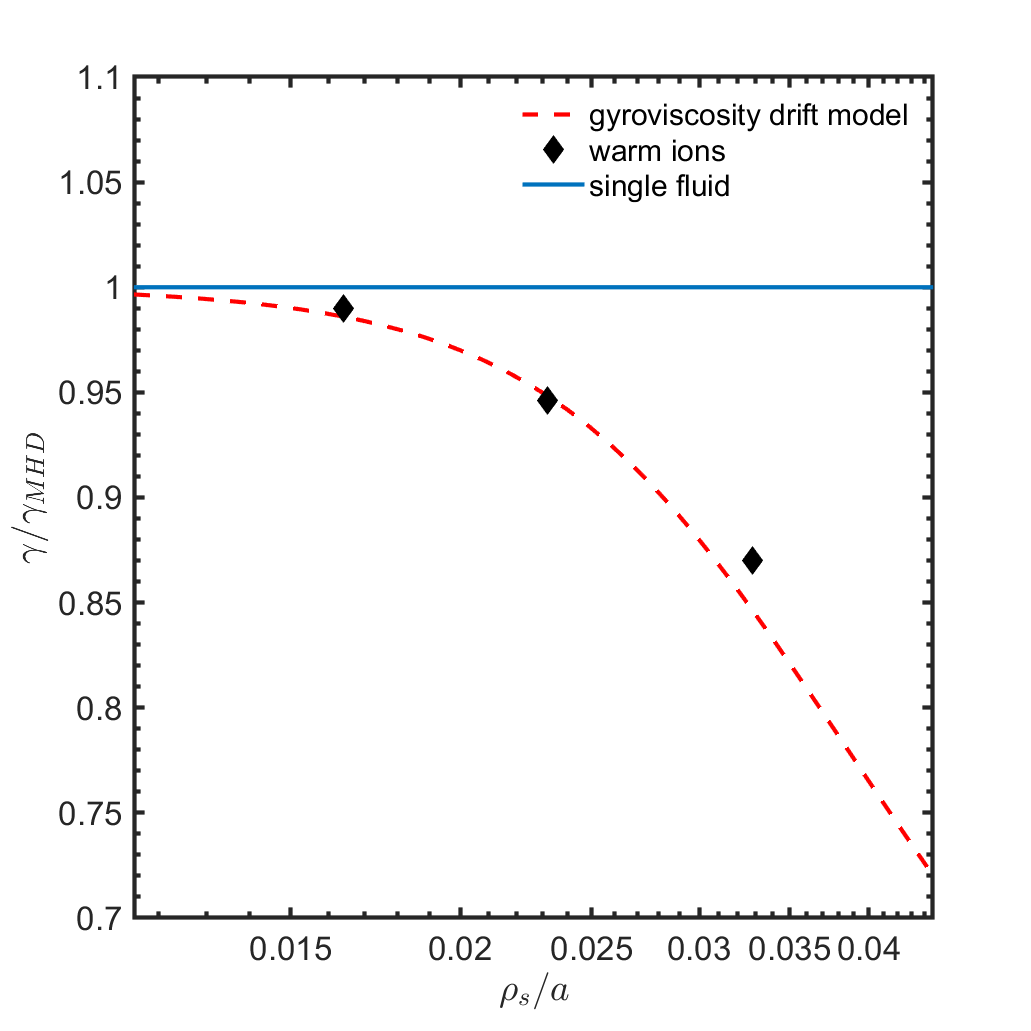}
\caption{Linear growth rates of $(-1,7)$ tearing mode under the parameter regime of the nonlinear simulation region.
} \label{fig:LinearGrowth}
\end{figure}

The linear growth rates of the dominant $(m,n)=(-1,7)$ tearing mode for the  setup  described above are illustrated  in Fig. \ref{fig:LinearGrowth}. In the limit of a small ion mass $m_i$, the extended MHD model converges to the single-fluid model. As $m_i$ increases into the drift-kinetic regime, which is relevant for the simulations performed, the tearing mode is stabilized by the warm-ion gyroviscous effect, leading to a decrease in the growth rate with increasing $m_i$. This result agrees well with the heuristic drift model proposed by King \cite{King-2012}, which is represented by the red dashed line in Fig. \ref{fig:LinearGrowth}.
As $m_i$ continues to increase into the kinetic Alfvén wave (KAW) mediated tearing regime, the ion gyroviscous radius becomes sufficiently large to decouple the tearing process from the ion fluid. Consequently, the two-fluid models exhibit enhanced growth rates, as previously reported by Mirnov et al. \cite{Mirnov-2004}. However, the corresponding parameters required for such decoupling significantly exceed the current operational range of the nonlinear simulations. 

A periodic cylinder with the KTX geometry parameters is employed, where the major radius $R=1.4\text{m}$ and the minor radius $a=0.38\text{m}$ \cite{Liu-2019}. A rectangular mesh composed of 32 by 64 (radial $\times$ toroidal) biquintic finite elements is used, along with $6$ Fourier components with $0\leq |m| \leq 5$ in the poloidal direction. The number of Fourier modes in the poloidal direction is optimized to minimize computational expense while maintaining sufficient spatial resolution in the toroidal direction.

The simulations start from a force-free equilibrium that satisfies the Ohmic steady-state condition, $\boldsymbol{E}_{eq} = -\boldsymbol{v}_{eq} \times \boldsymbol{B}_{eq} + \eta \boldsymbol{J}_{eq} = E_{loop} \boldsymbol{e}_\varphi$, where a constant toroidal electric field $E_{loop}$ is applied at the boundary as the loop voltage. The initial perturbation is set with a relative amplitude of $\delta B/B = 10^{-5}$. It is important to note that the initial equilibrium is one-dimensional and thus lacks a reversal at the boundary as per Cowling's theorem \cite{COWLING-1957}. At the radial boundary, the perturbed magnetic and velocity fields adhere to ideal and solid-wall no-slip conditions, whereas the temperature perturbation complies with the Dirichlet condition.

\section{Two-Fluid Effects on Sawtooth Behavior}

\begin{figure}[htbp]
\centering
\includegraphics[width=0.7\textwidth]{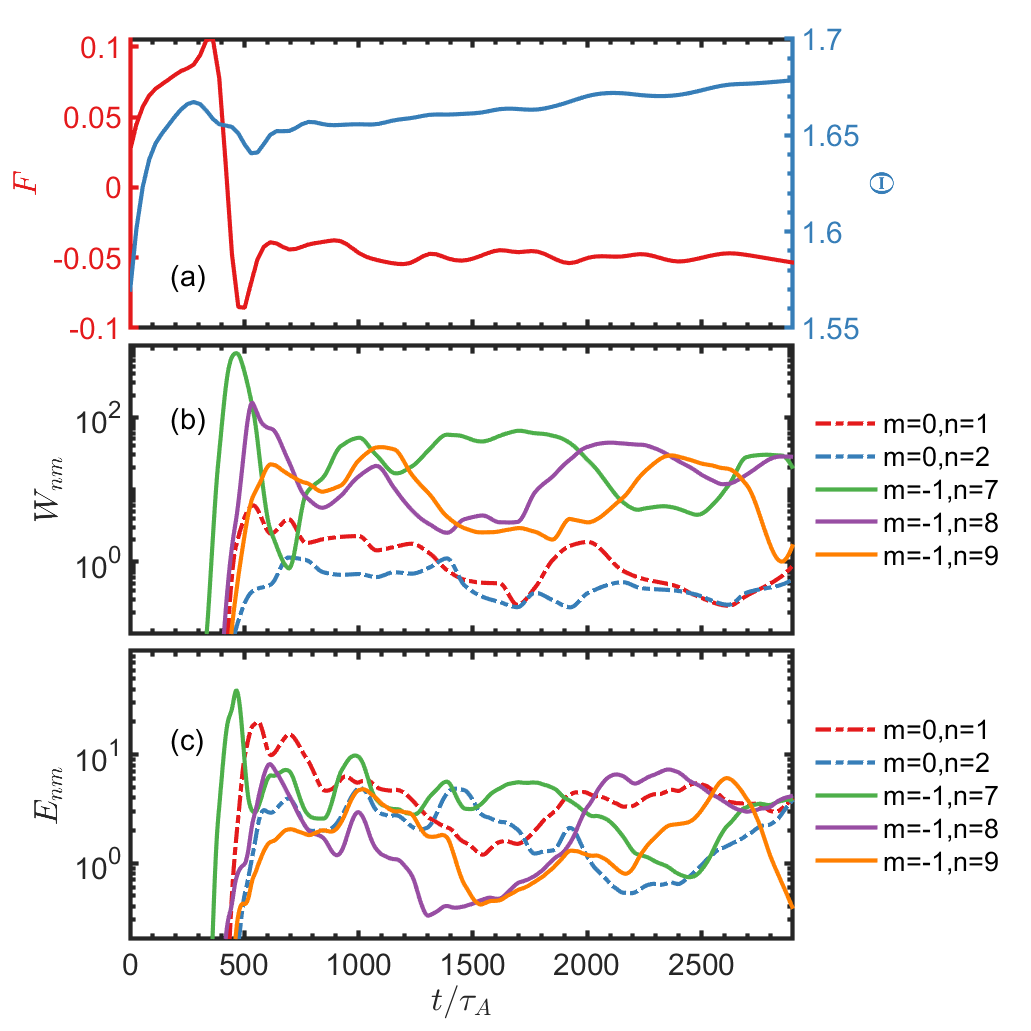}
\caption{Time evolution in the single-fluid simulation: (a) field-reversal parameter $F$ and pinch parameter $\Theta$, (b) magnetic energies $W_{mn}$, and (c) kinetic energies $E_{mn}$ of selected $(m,n)$ mode components.}
\label{fig:MHDEvolution}
\end{figure}

The results of single-fluid simulations in absence of  two-fluid effects are shown in Fig. \ref{fig:MHDEvolution}. Following an initial sawtooth relaxation event, the system self-organizes from a paramagnetic equilibrium state to a low-energy, sawtooth-free shallow reversal state. This transition  is manifest  in the temporal evolution of the reversal parameter $F$ and the pinch parameter $\Theta$, defined as $F = B_\varphi(a)/\langle B_\varphi \rangle$ and $\Theta = B_\theta(a)/\langle B_\varphi \rangle $, respectively. In the sawtooth-free state, the time-averaged reversal and pinch parameters are $\langle F \rangle = -0.05$ and $\langle \Theta \rangle = 1.68$. The toroidal reversal is maintained quasi-linearly by the $m = -1$ tearing modes. The amplitudes of the tearing modes with poloidal mode numbers $m = -1$ and toroidal mode numbers $n = 7, 8, 9$ exhibit comparable amplitudes and temporal fluctuations, indicating that the plasma is in an axisymmetric multi-helicity (MH) state. Furthermore, in the absence of sawtooth processes, the amplitude of the nonlinear $m = 0$ mode diminishes significantly.

\begin{figure}[htbp]
\centering
\includegraphics[width=0.7\textwidth]{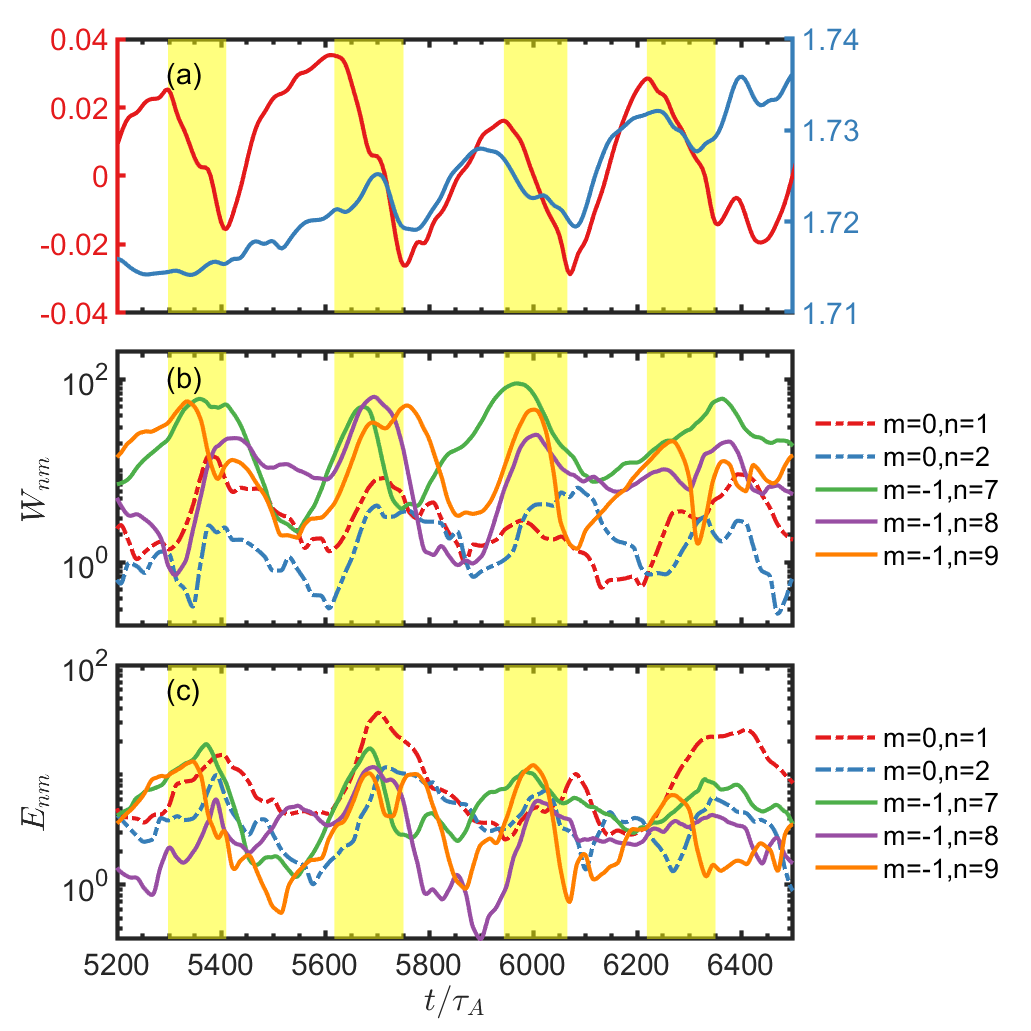}
\caption{Time evolution in the two-fluid simulation with $m_i/m_{i0}=1$: (a) field-reversal parameter $F$ and pinch parameter $\Theta$, (b) magnetic energies $W_{mn}$, and (c) kinetic energies $E_{mn}$ of selected $(m,n)$ mode components. The shaded intervals denote the phases of sawtooth events.}
\label{fig:2flEvolution}
\end{figure}

In contrast, with identical $S$ and $M$ parameters, a two-fluid MHD simulation with  an ion mass ratio of $m_i/m_{i0}=1$ exhibits distinct sawtooth cycles, as shown in Fig. \ref{fig:2flEvolution}. The  slow ramps in the reversal parameter $F$  are followed by rapid crashes, which are temporally coincident with bursts in the modal magnetic energy $W_{mn}$ and kinetic energy $E_{mn}$ (Fig. \ref{fig:2flEvolution} (b) and (c)), particularly for the nonlinear $m = 0$ mode. The plasma stays in the quasi-periodical sawtooth cycle  rather than relaxing to a relatively low-energy state. Since external drive is not fully balanced by the plasma's self-organization, the current profile  peaks again as the RFP plasma  gradually recovers from the deep reversal state. The free energy provided by the current gradient excites instabilities, leading to subsequent sawtooth crashes and rapid decreases in the $F$ parameter.

\begin{figure}[htbp]
\centering
\includegraphics[width=\textwidth]{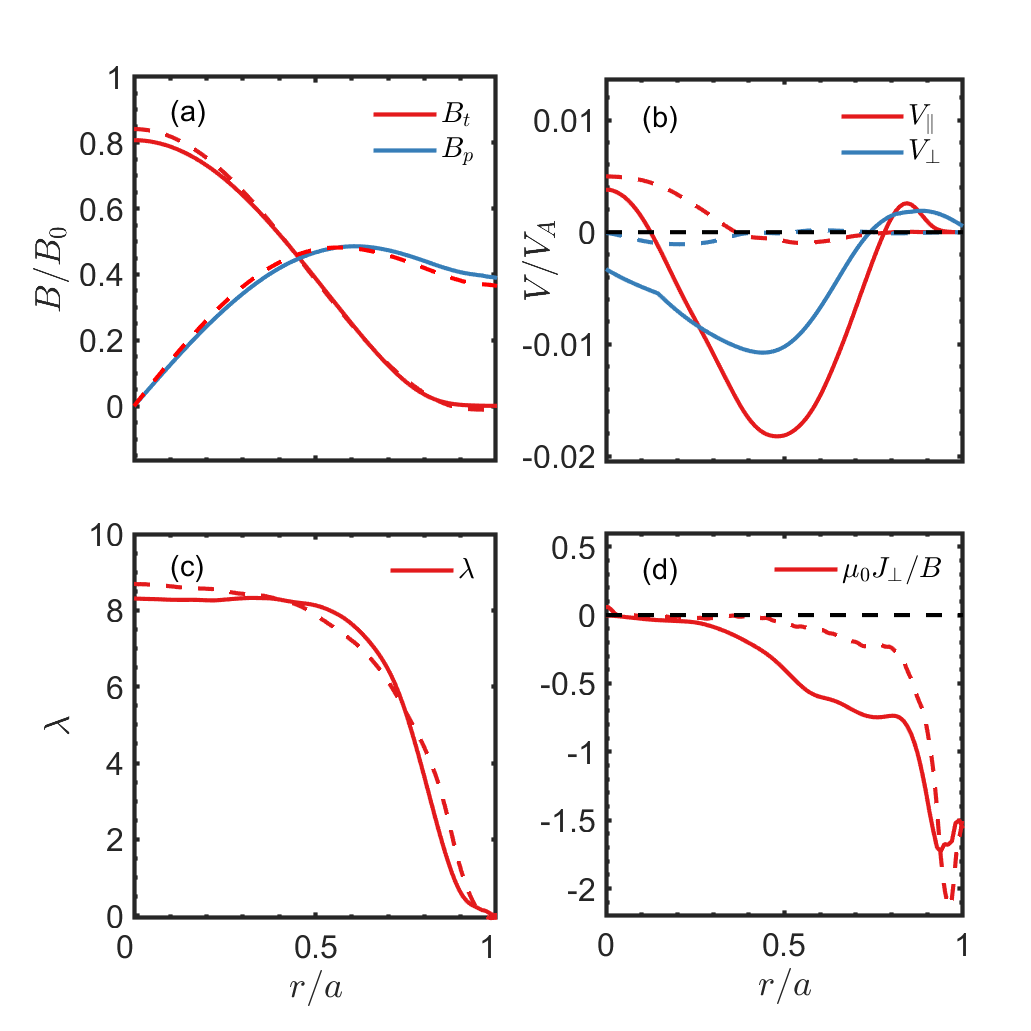}
\caption{ (a) The toroidal and poloidal magnetic field, (b) the parallel and perpendicular  velocity profiles, (c)  the normalized parallel and (d) perpendicular current profile. The solid line and the dashed line represent the results from the two-fluid model and the single-fluid model, respectively. All results are the outcome of long-time averaging.  }  \label{fig:Lambda}
\end{figure}

The presence of the Hall effect is known to  alter not only the sawtooth behavior but also the current and flow distributions in the relaxed state \cite{Lingam-2016,Gupta-2023}.  As shown in Fig. \ref{fig:Lambda}, two-fluid modifications to the relaxed state in our simulations yield weak variations in the time-averaged reversal and pinch parameters; $\langle F \rangle$ increases slightly from $-0.05$ to $0.01$, and $\langle \Theta \rangle$ increases from $1.68$ to $1.72$. The normalized parallel current profiles in the two-fluid case are similar to those of the single-fluid case. In contrast, the two-fluid case exhibits significant perpendicular flow, consistent with theoretical predictions \cite{Lingam-2016}, whereas the enhanced parallel flow is related to parallel momentum transport during sawtooth relaxation dynamics \cite{Sauppe-2017}. Additionally, in the two-fluid case, an extra perpendicular current arises from the perpendicular flow to maintain force balance, further affecting the magnetic field profile and the $\Theta$ parameter. However, due to the relatively weak two-fluid effects employed in the case, the slight deviation of equilibrium from the single-fluid case  is not the primary driver of the sawtooth behavior in the two-fluid simulations.

\begin{figure}[htbp]
\centering
\includegraphics[width=1.1\textwidth]{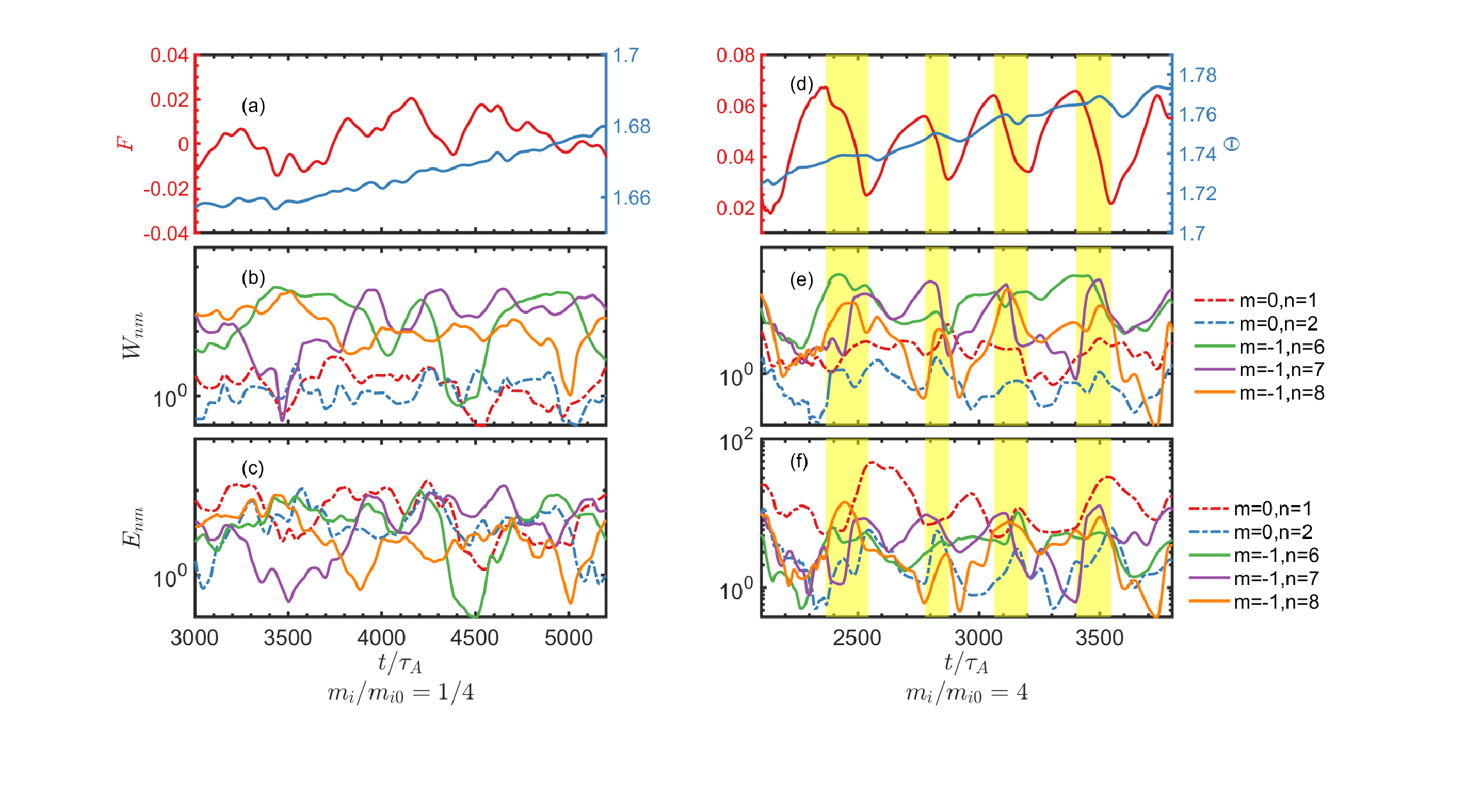}
\caption{ Left column: (a) Field reversal parameter $F$ and pinch parameter $\Theta$, (b) magnetic energies, and (c) kinetic energies of various $(m,n)$ mode components as functions of time in the two-fluid simulation with reduced ion mass $m_i/m_{i0}=1/4$. Right column: (d) Field reversal parameter $F$ and pinch parameter $\Theta$, (e) magnetic energies, and (f) kinetic energies of various $(m,n)$ mode components as functions of time in the two-fluid simulation  with increased ion mass $m_i/m_{i0}=4$.}  \label{fig:2flEvolution2}
\end{figure}

The direct comparison of two-fluid and single-fluid cases under identical $S$ and $M$ parameters clearly demonstrates that two-fluid effects enhance the occurrence of sawtooth activity. This conclusion is further supported by comparative simulations that alter the strength of two-fluid effects. For example, in the two-fluid case with a reduced ion mass ratio of $m_i/m_{i0}=1/4$ (Fig. \ref{fig:2flEvolution2}), the amplitudes of bursts in $W_{mn}$ and $E_{mn}$ decrease, and the characteristic sawtooth ramps and crashes of $F$ and $\Theta$ largely disappear. By contrast, in the two-fluid case with an increased ion mass ratio of $m_i/m_{i0}=4$ (Fig. \ref{fig:2flEvolution2}), the system exhibits similar but more pronounced sawtooth behavior. Given the results in Fig. \ref{fig:LinearGrowth}, which shows that two-fluid effects suppress the linear growth rate of tearing modes within the considered parameter range, we may  hypothesize that two-fluid effects promote the onset of sawtooth behavior through  nonlinear mechanism.

\begin{figure}[htbp]
\centering
\includegraphics[width=1.1\textwidth]{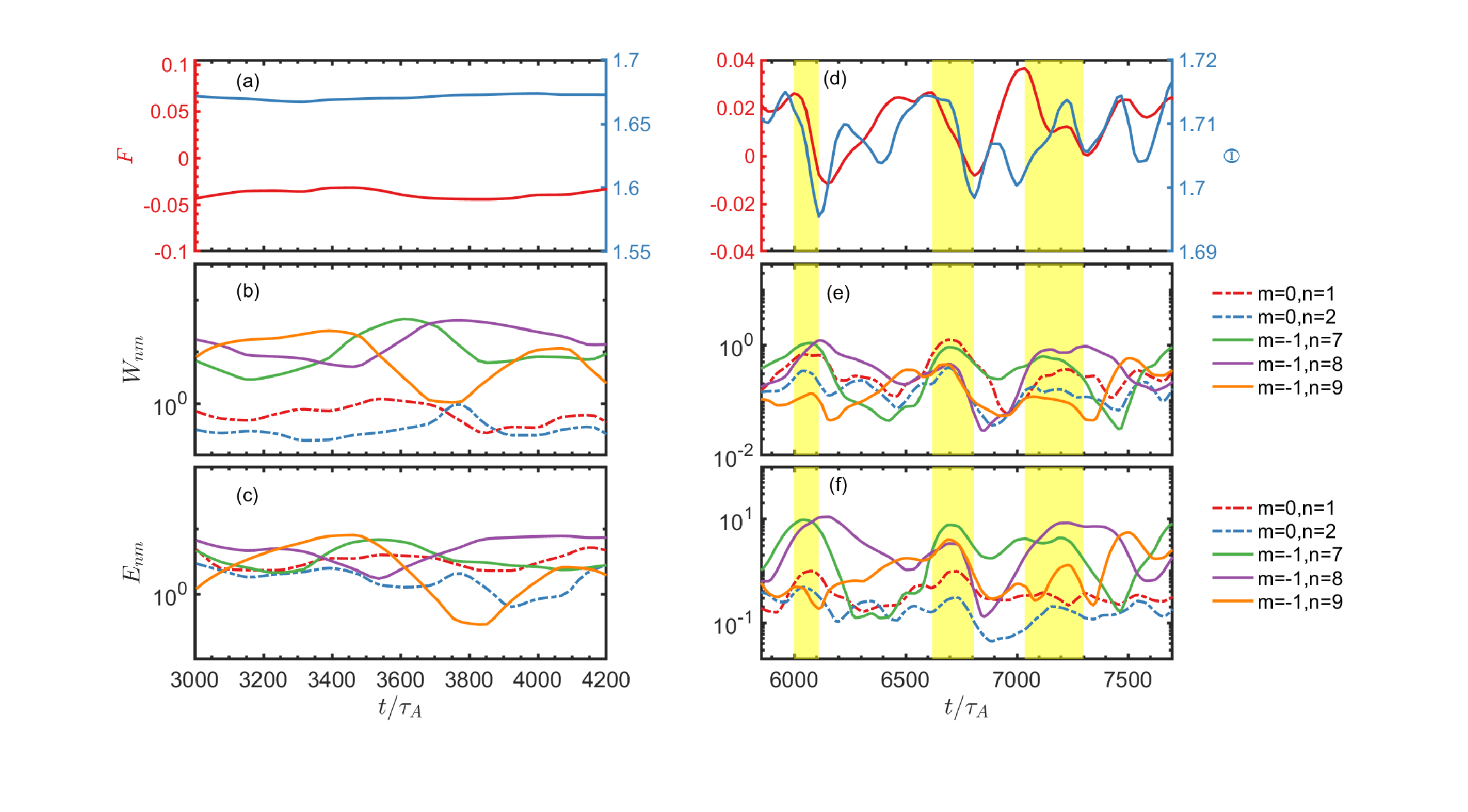}
\caption{ Left column: (a) Field reversal parameter $F$ and pinch parameter $\Theta$, (b) magnetic energies, and (c) kinetic energies of various $(m,n)$ mode components as functions of time in the single-fluid simulation  with anisotropic thermal conductivity. Right column: (d) Field reversal parameter $F$ and pinch parameter $\Theta$, (e) magnetic energies, and (f) kinetic energies of various $(m,n)$ mode components as functions of time in the two-fluid simulation with  anisotropic thermal conductivity.}  \label{fig:AnisolEvolution}
\end{figure}

\begin{figure}[htbp]
\centering
\includegraphics[width=0.8\textwidth]{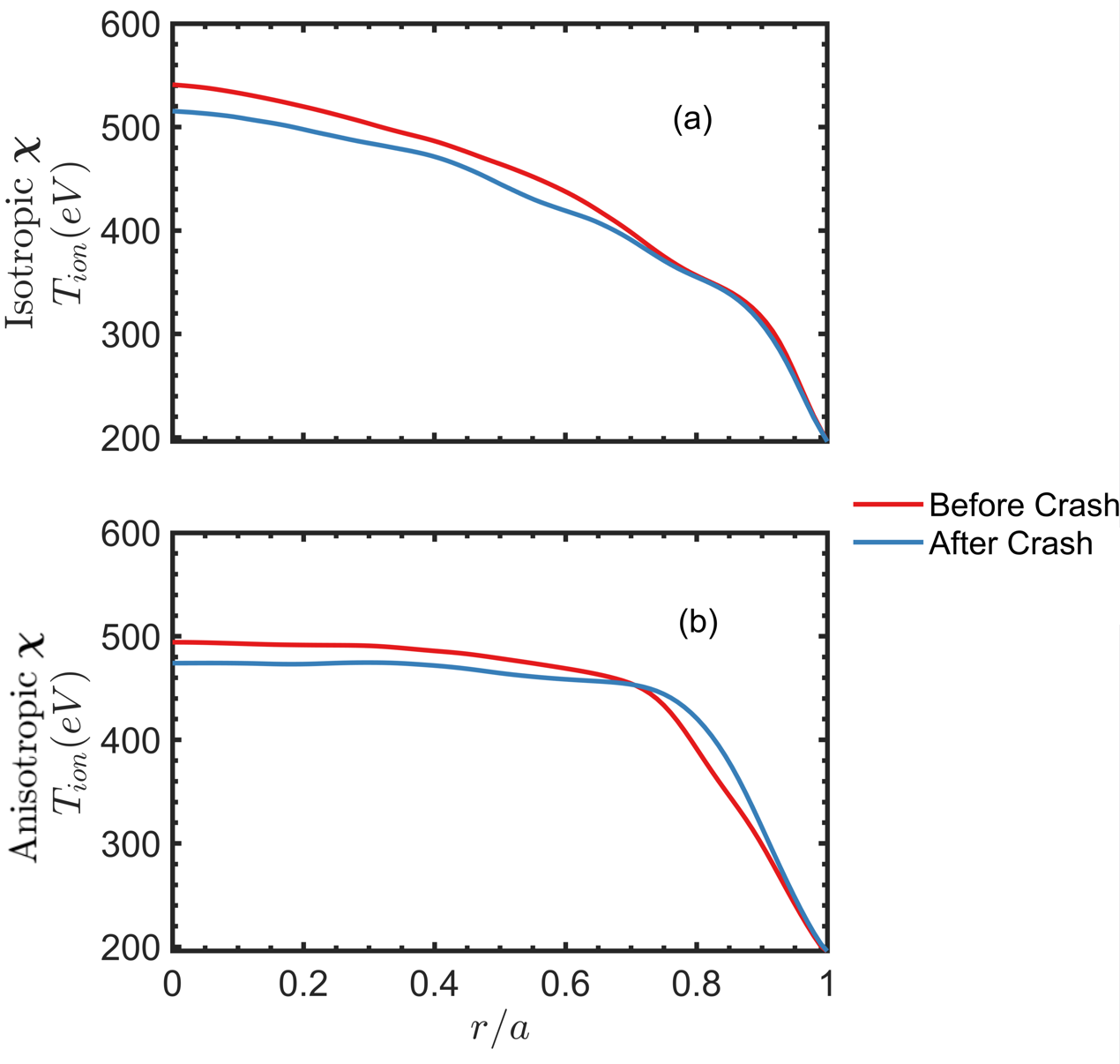}
\caption{ The radial profiles of ion temperature before and after the sawtooth crash, obtained from two-fluid simulations with (a) isotropic thermal conductivity and (b) anisotropic thermal conductivity.}  \label{fig:RadialTion}
\end{figure}

The simulations presented in this work primarily employ an isotropic thermal conductivity model, with both the parallel and perpendicular conductivities set to $0.036\,\mathrm{m^2/s}$. To assess the influence of anisotropic heat transport, additional comparison simulations have been performed using an anisotropic thermal conductivity model, where the parallel conductivity is increased by a factor of $10^6$ to $3.6\times10^{4}\,\mathrm{m^2/s}$ while the perpendicular conductivity remains unchanged.
As shown in Fig.~\ref{fig:AnisolEvolution}, the sawtooth-free state remains in the single-fluid simulation with anisotropic thermal conductivity, consistent with the isotropic result. In the two-fluid anisotropic case, sawtooth oscillations persist as well. The corresponding radial profiles of ion temperature before and after the sawtooth crash are shown in Fig.~\ref{fig:RadialTion}. With anisotropic thermal conductivity, the core temperature gradient is significantly reduced owing to the greatly enhanced parallel heat transport along the magnetic field lines. By contrast, the edge temperature gradient remains relatively well preserved, as the good magnetic flux surfaces in this region effectively inhibit radial heat loss. Despite these differences, both the isotropic and anisotropic cases exhibit a pronounced drop in core temperature during the sawtooth crash, which suggests they share the common path of  rapid release of thermal energy through magnetic reconnection.

\section{The flux-pumping Mechanism in RFP}

To quantitatively describe the two-fluid effect on the sawtooth process, we analyze the conditions required for the plasma to maintain a sawtooth-free state \cite{Jardin-2020,McCollam-2026}. This analysis centers on the  self-organization mechanism  of magnetic flux-pumping,  where saturated MHD instabilities generate a dynamo electric field that redistributes magnetic flux and current radially. Starting from the time evolution equation of the magnetic vector potential $\boldsymbol{A}$ in the Coulomb gauge:
\begin{equation}
\frac{\partial \boldsymbol{A}}{\partial t} = -\boldsymbol{E} - \nabla\Phi
\end{equation}
Substituting Ohm's law and   neglecting the small diamagnetic term, and projecting along the magnetic field direction, we obtain the following form of the magnetic flux evolution equation:
\begin{equation} 
-\frac{\partial(A^{\text{plasma}}_{\parallel}+A^{\text{loop}}_{\parallel})}{\partial t} = \eta J_{\parallel} - \boldsymbol{e}_b\cdot(\boldsymbol{v}\times\boldsymbol{B}) + \boldsymbol{e}_b\cdot \left ( \frac{\boldsymbol{J}\times\boldsymbol{B}}{ne} \right ) + \boldsymbol{e}_b\cdot\nabla\Phi\label{eqn:FluxEvo}	
\end{equation}
where $A^{\text{plasma}}_{\parallel}$ and $A^{\text{loop}}_{\parallel}$ are the parallel vector potential components generated by the plasma and the external ohmic drive, respectively, $\Phi$ represents the electrostatic potential, and other symbols are conventional. After performing a surface average $\langle f\rangle=\iint{}d\theta{}d\varphi{}f/(4\pi^2)$ and keeping only the zeroth-order components, the electrostatic potential term vanishes. Near the rational surface, the helical magnetic field 
\begin{equation}
\boldsymbol{B}_*=\nabla \Psi_h\times \nabla\zeta
=\nabla\times(\Psi_h\nabla\zeta)
\approx\nabla \times(A_\parallel\boldsymbol{e}_b)
\end{equation}
where $\Psi_h$ is the helical flux function and $\zeta$ is the helical coordinate along the field line. Hence Eq. (\ref{eqn:FluxEvo}) can be written as the evolution equation for the helical flux function $\Psi_h$ as:

\begin{equation}
\frac{\partial \Psi_h}{\partial t} = -\langle \eta J_\parallel \rangle + \langle\boldsymbol{v}\times\boldsymbol{B}\rangle_\parallel - \left\langle\frac{\boldsymbol{J}\times\boldsymbol{B}}{ne}\right\rangle_\parallel + E^{loop}_\parallel
\end{equation}

Across all cases, the resistive term $\langle \eta J_{\parallel} \rangle$  provides  the dominant term in the parallel electric field $E_{\parallel}$ over the radius. The   fluctuation induced dynamo electric field, comprising the MHD dynamo   $\langle\boldsymbol{v}\times\boldsymbol{B}\rangle_\parallel$  and Hall dynamo term $ - \langle{\boldsymbol{J}\times\boldsymbol{B}}/{ne}\rangle_\parallel$,  reflects the self-organization behavior of the plasma, with contributions from multiple modes responsible for magnetic flux-pumping.  For the plasma to achieve and sustain a steady magnetic state, the net parallel  electric field $E_{\parallel}-E_{\parallel}^{\mathrm{loop}}$  must vanish.  The deviation between  $E_{\parallel}$ and  $E_{\parallel}^{\mathrm{loop}}$ indicates finite flux transport and an evolving magnetic field.

In the sawtooth-free regime, current profiles are broadened by a quasi-linear MHD dynamo in either a quasi-single helicity (QSH) or MH state—driving the system toward a relaxed, reversed-field configuration. This dynamo mechanism is analogous to magnetic flux-pumping in tokamaks, where a saturated $(m,n)=(-1,1)$ quasi-interchange instability generates an effective negative loop voltage in the plasma core. This process flattens the current density profile and maintains the core safety factor, $q_0$, above unity, thereby preventing sawteeth \cite{Jardin-2015,Krebs-2017,Zhang-2025}. Unlike tokamak plasmas, which are characterized by a strong toroidal field and a weak poloidal field, RFPs exhibit comparable toroidal and poloidal field magnitudes. This results in a safety factor, $q$, that is well below unity. Consequently, $m=-1$ tearing modes predominantly govern the stability and self-organization of RFP plasmas. Unlike the QSH state, where a single dominant mode acts similarly to the $(1,1)$ quasi-interchange mode in tokamaks, the MH state considered in this work involves complex mode interactions that fundamentally alter the flux-pumping dynamics.

\begin{figure}[htbp]
\centering
\includegraphics[width=\textwidth]{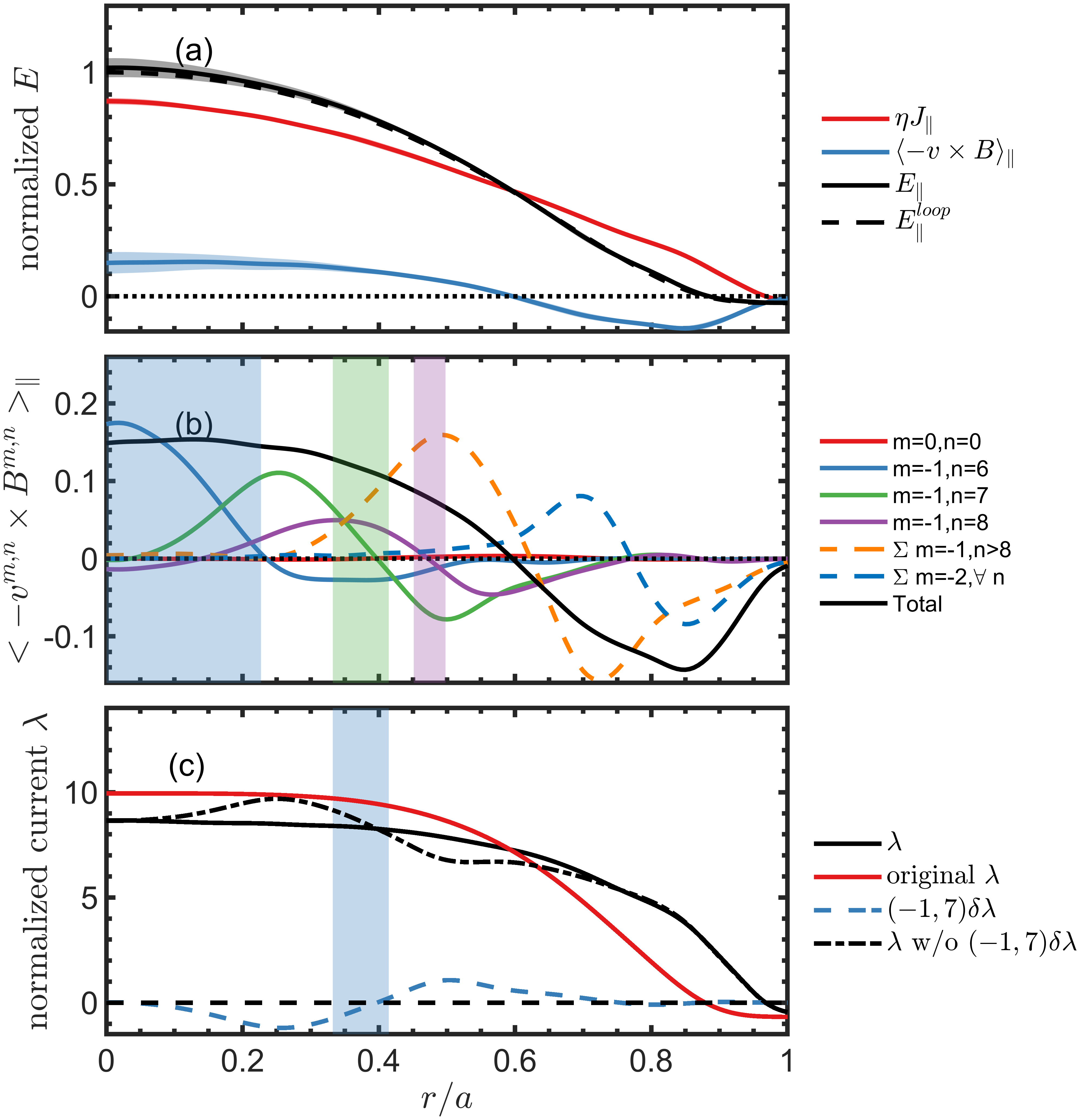}
\caption{ (a) The parallel components   in the generalized Ohm's equation of the sawtooth-free single-fluid case, where lines and shaded regions
represent the average value and the range of  electric field at various times, respectively. Modal decomposition of  (b) MHD dynamo electric fields. (c) The normalized current profile with and without the modification of saturated  $(-1,7)$ tearing mode.}  \label{fig:MHDDynamo}
\end{figure}

In the single-fluid case illustrated in Fig. \ref{fig:MHDDynamo},  the parallel electric fields is normalized by the applied loop electric field $E^{\text{loop}}$. The MHD dynamo is positive in the core  and  negative outside. This balances the difference between the resistive term $\eta J_{\parallel}$ and $E_{\parallel}^{\text{loop}}$, facilitating the outward flux-pumping needed to maintain magnetic relaxation. The dominant contributions  to the averaged parallel dynamo electric field $\langle -\boldsymbol{v}\times\boldsymbol{B} \rangle_{\parallel}$ come from  tearing modes with $m = -1$ and $n = 6, 7, 8$ as shown in Fig. \ref{fig:MHDDynamo}, panel (b). The peaks of these modes are located slightly outside their respective rational surfaces. Similarly, for higher $n$ modes, the peaks in the dynamo electric field shift sequentially  outward from the interior toward the edge.  The amplitude of the $m=0$ component is negligible compared to that of the contributing $m=-1$ modes. 

Previous theoretical studies have highlighted the point that saturated  tearing modes can lead to a dynamo effect that reduces the current gradient near their resonant surfaces \cite{Luo-2021,Luo-2022}. This offers a complementary perspective on the dynamo action, as illustrated in Fig. \ref{fig:MHDDynamo}, panel (c). Here, the original normalized current $\lambda$, which represents the current profile in the absence of nonlinear dynamo modification (corresponding to the initial paramagnetic pinch state), decreases sharply  from the core to the edge. In contrast, the actual $\lambda$ profile in the sawtooth-free relaxed state is relatively flat. Comparison with the dashed line, which represents the current profile without the $(-1,7)\ \delta\lambda$ contribution, shows that the dynamo effect from saturated tearing modes flattens the current profile, consistent with theoretical predictions.  Consequently, this dynamo action acts as a self-regulating mechanism that reduces current gradients, leading to a more relaxed plasma state.

The dynamo mechanism in RFP plasmas demonstrated above  differs  from  the flux-pumping in tokamaks. In a tokamak, the dynamo effect is typically driven  by $(1,1)$ interchange mode to maintain the central safety factor $q_0 > 1$ and thereby prevent sawtooth oscillations \cite{Yu-2024,Jardin-2020}. In contrast, the parallel dynamo electric field in the MH state of an RFP is sustained by multiple saturated $m=-1$ tearing modes. This  dynamo electric field transports magnetic flux radially, converting energy from the externally  applied  toroidal ohmic drive into  poloidal current, enabled by the strong magnetic shear characteristic of RFPs. Overall, the RFP dynamo is  global and strong over  the radius, in  contrast to the localized nature of its tokamak counterpart.  Within the single-fluid MHD framework, this dynamo process effectively maintains a stable, sawtooth-free magnetic equilibrium. The following  section will demonstrate how the inclusion of two-fluid effects can interrupt this steady state.

\section{Two-Fluid Effects on Dynamo Electric Fields}

\subsection{Dynamo Electric-Field Profiles}

\begin{figure}[htbp]
\centering
\includegraphics[width=\textwidth]{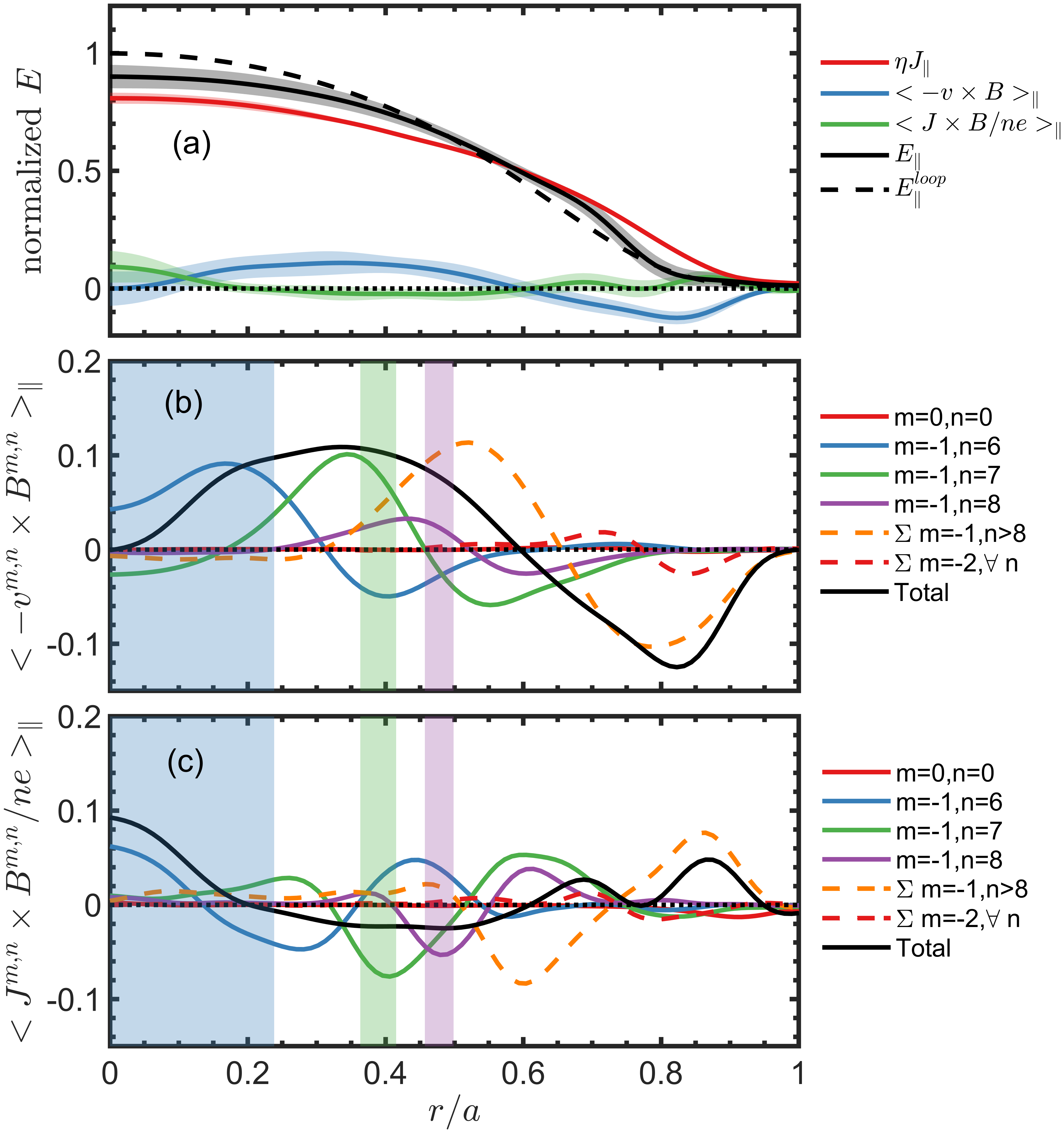}
\caption{ (a) The parallel components   in the generalized Ohm's equation during the rising phase of the sawtooth cycle, where lines and shaded regions
represent the average value and the range of  electric field at various times, respectively. Modal decomposition of  (b) MHD dynamo and (c) Hall  dynamo electric fields.}  \label{fig:2flDynamo}
\end{figure}

\begin{figure}[htbp]
\centering
\includegraphics[width=\textwidth]{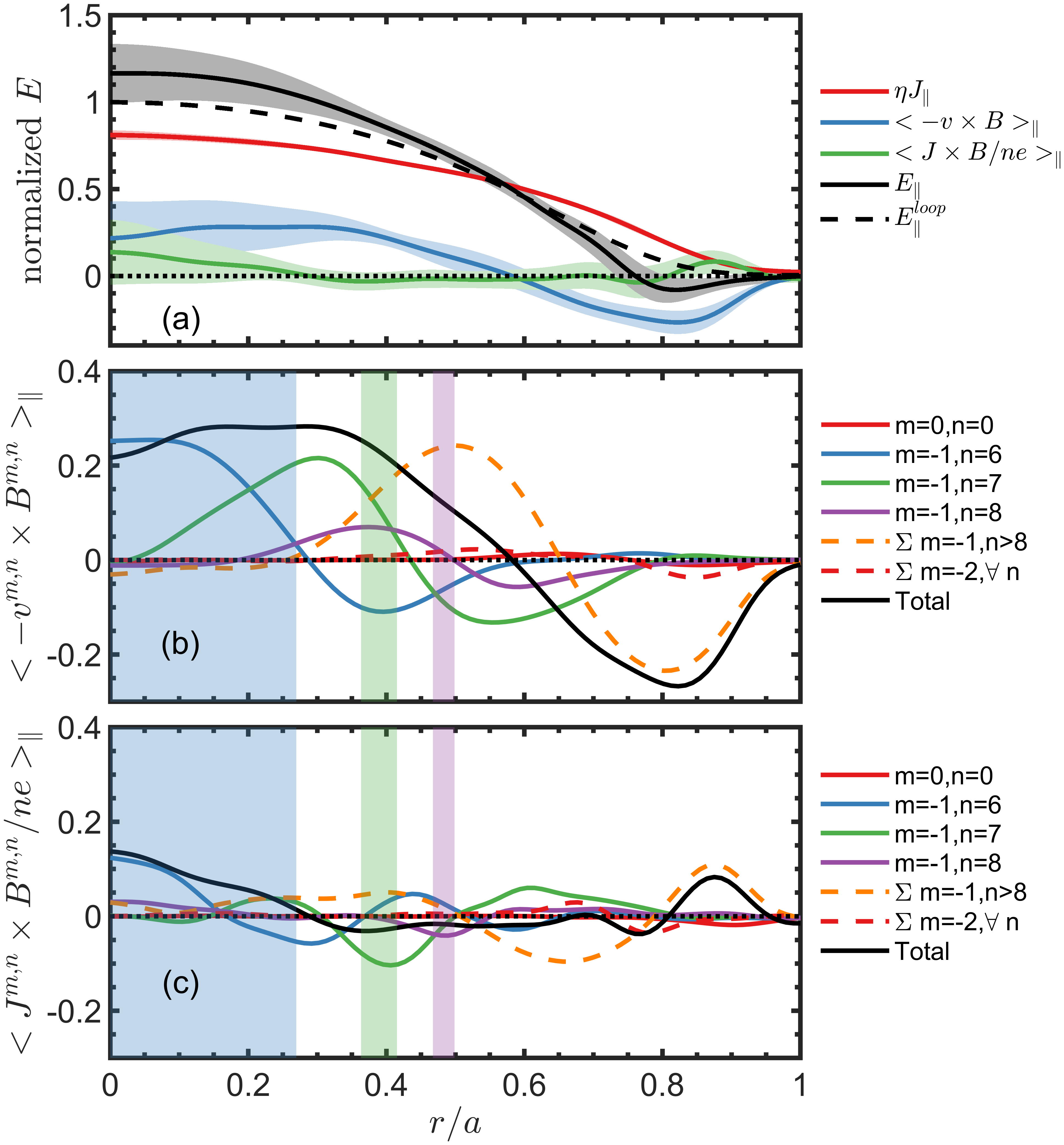}
\caption{ (a) The parallel components   in the generalized Ohm's equation during the crash phase of the sawtooth cycle, where lines and shaded regions
represent the average value and the range of  electric field at various times, respectively. Modal decomposition of  (b) MHD dynamo and (c) Hall  dynamo electric fields.}  \label{fig:2flDynamo2}
\end{figure}

Inclusion of two-fluid effects in simulations enables the  quasi-periodic sawtooth behavior, as illustrated in Figs.\ref{fig:2flDynamo} and \ref{fig:2flDynamo2}. Fig. \ref{fig:2flDynamo} presents the parallel electric-field components during the rising phase, along with the individual contributions of various modes to the MHD and Hall dynamo fields. 
A nonzero net electric field drives temporal variations in the magnetic field, which are negative in the core region and positive outward of the radius $r/a = 0.50$. This profile indicates that the dynamo amplitude is insufficient to sustain the magnetic configuration in a fully relaxed state. Consequently, the externally applied loop voltage continuously injects  energy into the system, transporting magnetic flux from the edge toward the core. The sum of the Hall and MHD dynamo fields maintains similar profile structures consistent with single-fluid results \cite{Sauppe-2017}, where the MHD dynamo is weakened but remains dominant, while the Hall term provides a significant contribution, particularly within the plasma core.

A mode-by-mode decomposition of the fluctuation-induced dynamo fields shows that, in the two-fluid simulations, the contributions from various tearing modes remain localized near their respective resonant surfaces. A primary distinction from single-fluid results is the radial outward shift of the electric field peaks due to two-fluid effects, while the locations of the resonant surfaces remain largely unchanged. Although the axisymmetric magnetic field evolves during nonlinear relaxation, the resulting rational-surface displacement is small and does not account for the observed peak shift. For instance, the $m=-1,n = 6, 7, 8$ modes peak at $r/a = 0.025, 0.26, 0.33$ in the single-fluid case (Fig. \ref{fig:MHDDynamo}), but shift to $r/a = 0.17, 0.34, 0.42$ in the two-fluid case. The Hall dynamo contributions from individual $m=-1$ modes are concentrated near their resonances and are characterized by a negative peak.

During the sawtooth crash phase, the net parallel electric field $E_{\parallel} - E^{loop}_{\parallel}$ reverses its sign to  become positive inside and negative outside, with an  amplitude  significantly larger than that during the rising phase. This reversal represents the rapid release of free energy through magnetic reconnection, which flattens the peaked current profile and returns the plasma to a relaxed state. Although the nonlinear $m=0$ mode is critical to the reconnection process, it does not contribute to the time-averaged dynamo electric field. Apart from  the increased magnitude, the profiles of the MHD and Hall dynamos during the crash are qualitatively similar to those in the rising phase.

\begin{figure}[htbp]
\centering
\includegraphics[width=\textwidth]{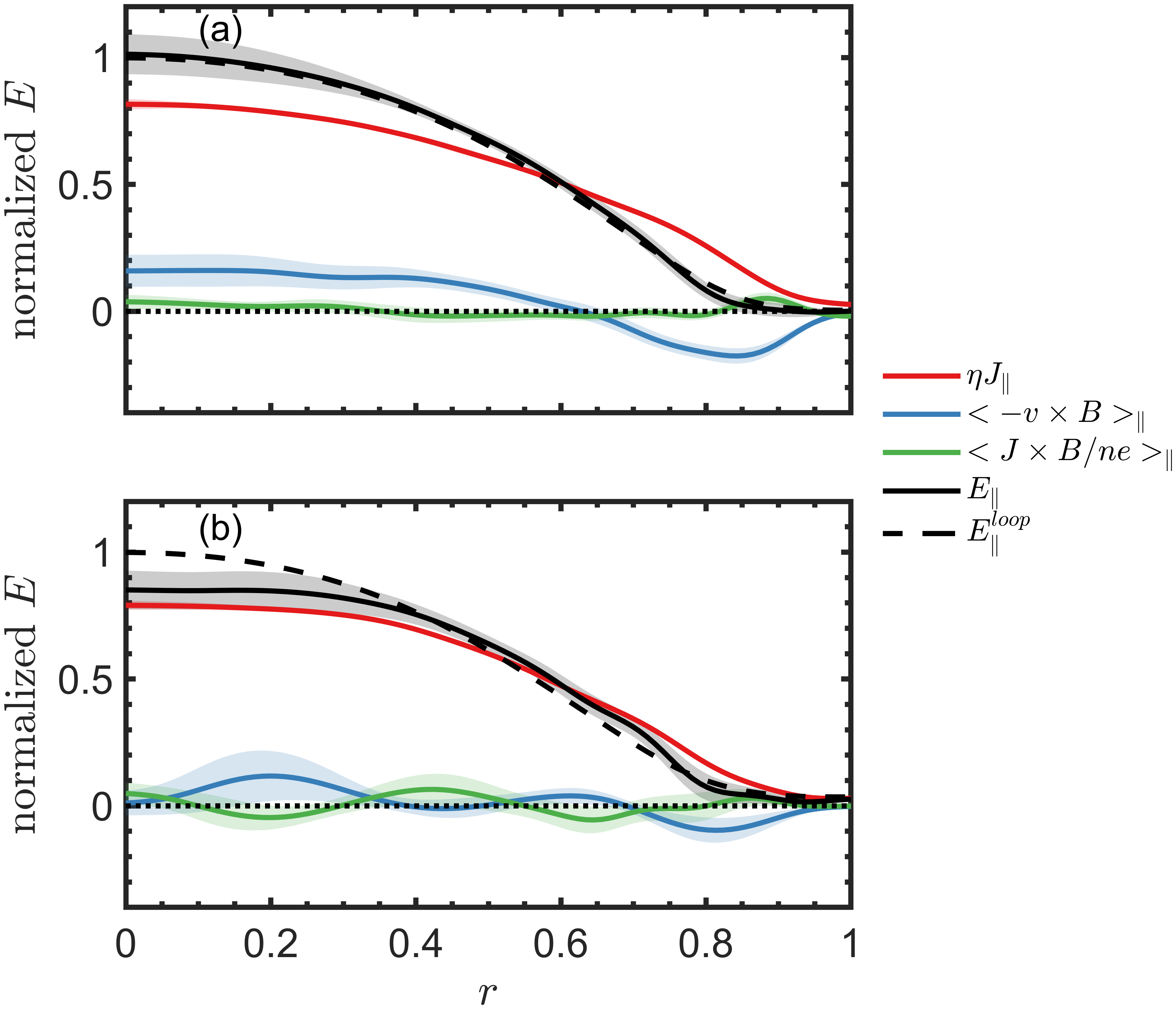}
\caption{ The parallel components   in the generalized Ohm's equation of the (a) reduced $m_i$ case and (b) increased $m_i$ case, where lines and shaded regions represent the average value and the range of  electric field at various times, respectively. }  \label{fig:VarMisDynamo}
\end{figure}

The influence of two-fluid effects is further investigated by comparing cases with varying ion mass as shown in Fig. \ref{fig:VarMisDynamo}.  With a reduced ion mass, the Hall dynamo $ -\langle{\boldsymbol{J}\times\boldsymbol{B}}/{ne}\rangle_\parallel$ weakens and becomes negligible. The remaining  results align closely with the single-fluid case and  sawtooth activity is mitigated.  Conversely, enhanced two-fluid effects  due to an increased  $m_i$ amplify the Hall dynamo amplitude while reducing the MHD dynamo amplitude, causing the two terms to become comparable in magnitude. Simultaneously, the net parallel dynamo field becomes larger, which leads to more frequent sawtooth relaxations.

\subsection{Tearing-Mode Structure and Helical Dynamo Contribution}

To investigate the origin of the tearing mode dynamo electric field in flux-pumping and the influence of two-fluid effects, the perturbed magnetic field $\tilde{\boldsymbol{b}}$, velocity $\tilde{\boldsymbol{v}}$, and current density $\tilde{\boldsymbol{J}}$ of the $(m,n)=(-1,7)$ mode are projected onto a helical coordinate system. Let ${\boldsymbol{e}}_r$ denote the direction normal to the helical surface, ${\boldsymbol{e}}_\zeta = (-{\boldsymbol{e}}_\varphi m/r+{\boldsymbol{e}}_\theta n/R_0)/l$, and ${\boldsymbol{e}}_\xi={\boldsymbol{e}}_\zeta\times{\boldsymbol{e}}_r = (- {\boldsymbol{e}}_\varphi n/R_0 - {\boldsymbol{e}}_\theta m/r)/l$, where $l=\sqrt{(m/r)^2+(n/R_0)^2}$ and $\xi=-m\theta-n\varphi$. Near the $(m,n)$ resonant surface, ${\boldsymbol{e}}_\zeta$ and ${\boldsymbol{e}}_\xi$ approximately represent the directions parallel and perpendicular to the equilibrium magnetic field, respectively.

\begin{figure}[htbp]
\centering
\includegraphics[width=.7\textwidth]{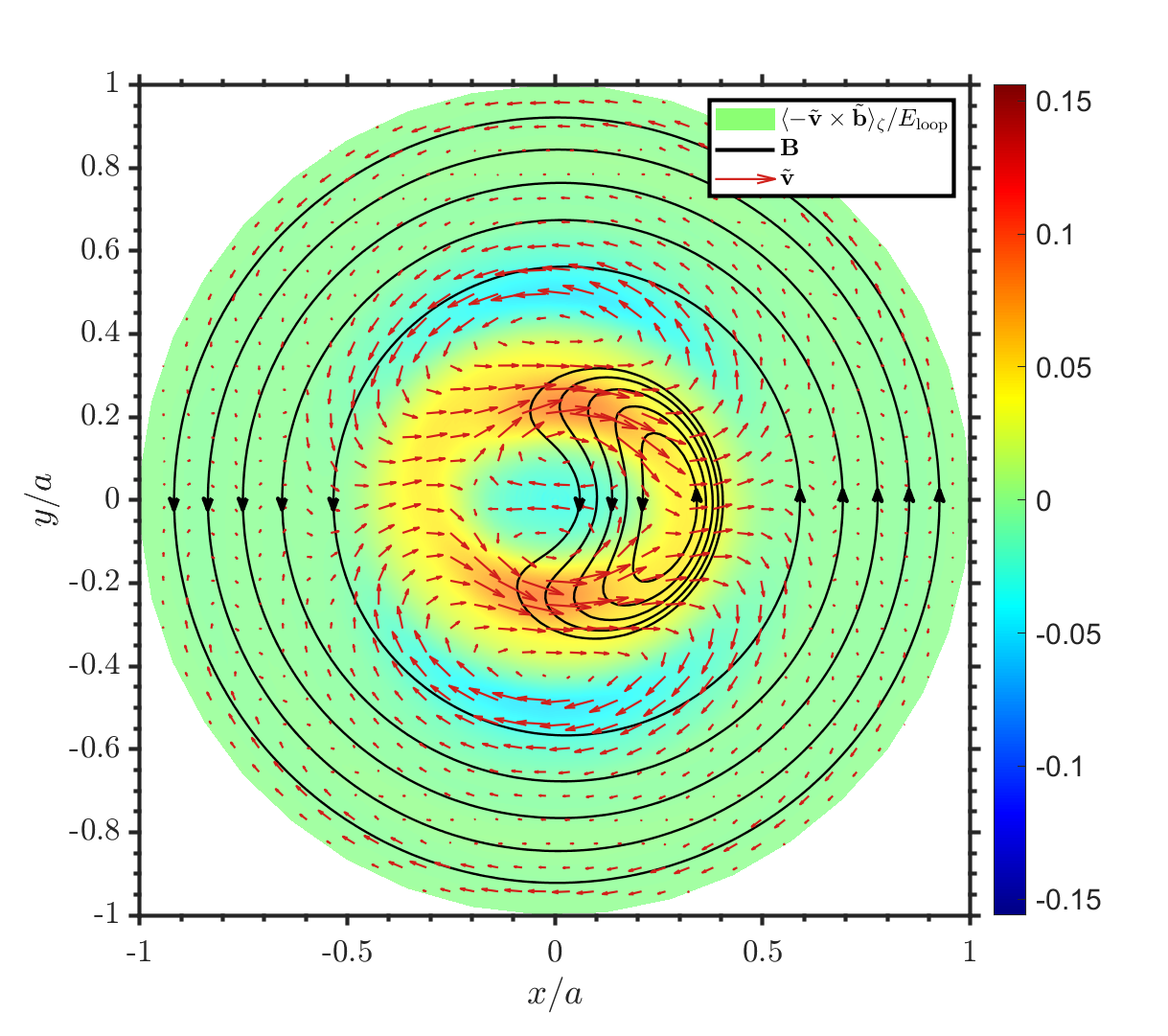}
\caption{ Helical projection of the $(-1,7)$ perturbed flow  $\boldsymbol{\tilde{v}}_\zeta $ (arrows) and the corresponding dynamo field in the parallel direction  $\langle -\boldsymbol{\tilde{v}}\times \boldsymbol{\tilde{b}}\rangle _\zeta$. The color scale represents the amplitude and direction of  the vector field component in $\zeta$ direction. Positive (red) and negative (blue) values denote outward and inward directions across the helical projection plane, which is aligned parallel and anti-parallel to the mean magnetic field. The equilibrium magnetic field is used only to outline the island structure; the dynamo electric field is computed from the perturbed magnetic field $\tilde{\boldsymbol{b}}$ alone.}  \label{fig:MHDTMContour}
\end{figure}

The parallel MHD dynamo  electric field and perpendicular flow for the single-fluid case are illustrated in Fig. \ref{fig:MHDTMContour}, where $\boldsymbol{e}_\zeta$ points out of the page and ${\boldsymbol{e}}_\xi$ is oriented counter-clockwise in the poloidal plane. For the $(-1,7)$ mode, the perpendicular plasma flow is characterized by two vertically symmetric vortices: plasma flows from the O-point to the X-point outside the resonant surface, and returns via the core to the O-point. The products  $\tilde{v}_{\xi} \tilde{b}_{r}$ and $-\tilde{v}_{r} \tilde{b}_{\xi}$ generate a dynamo electric field that is positive inside and negative outside the resonant surface, as shown in Fig. \ref{fig:MHDDynamo}.

\begin{figure}[htbp]
	\centering
	\subfloat[]{\includegraphics[width=.45\textwidth]{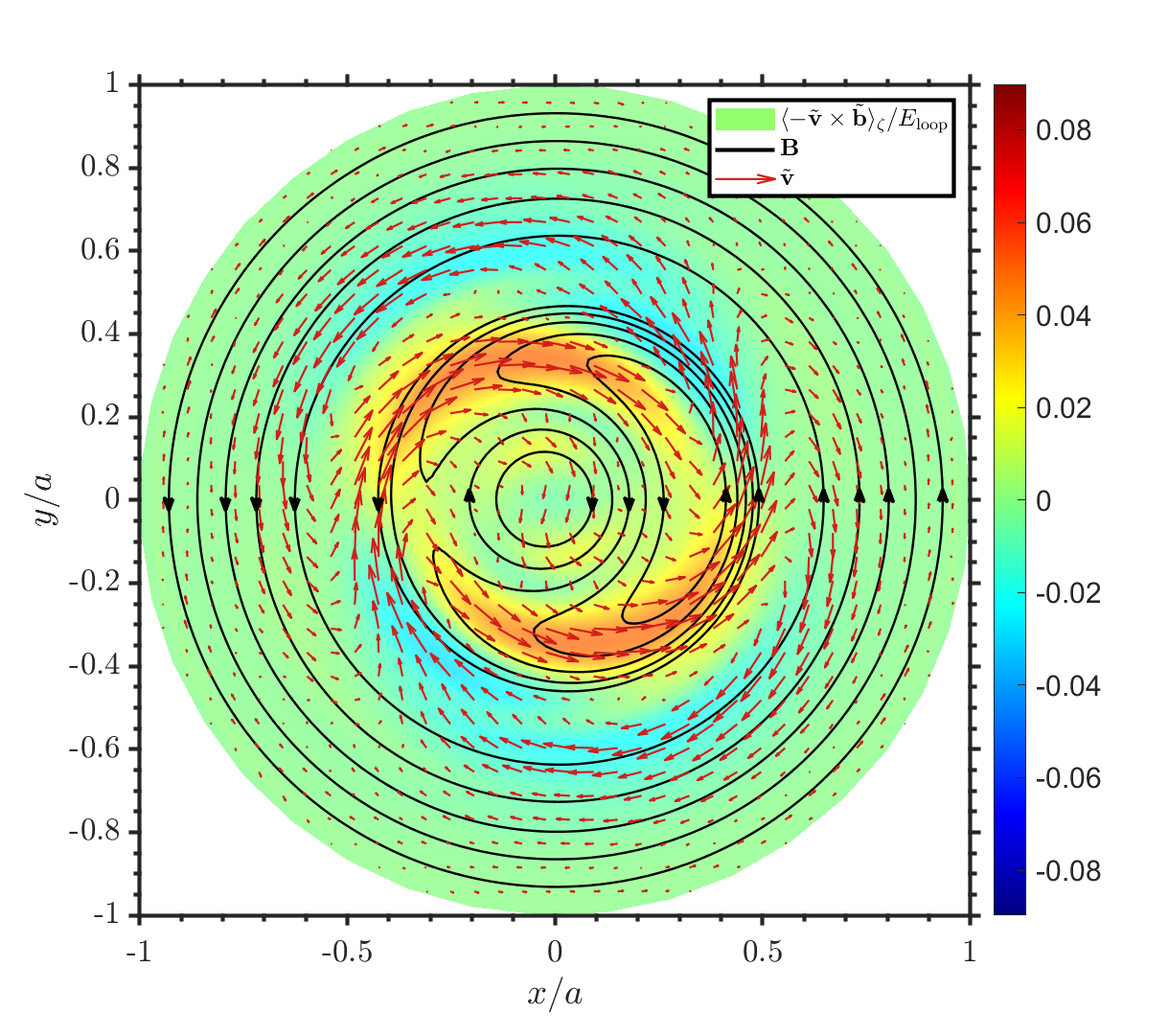}}\hspace{5pt}
	\subfloat[]{\includegraphics[width=.45\textwidth]{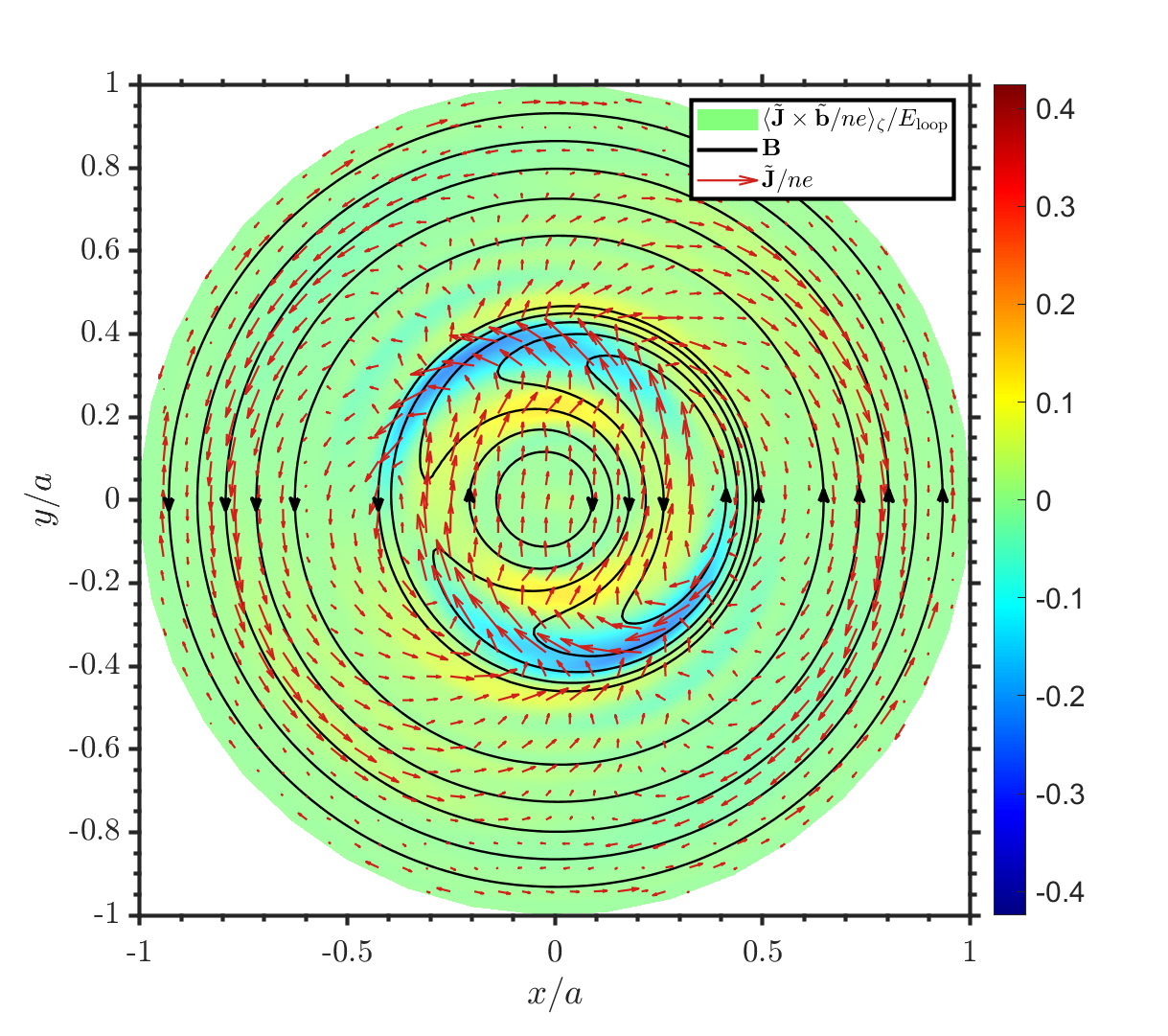}}\\
	\subfloat[]{\includegraphics[width=.45\textwidth]{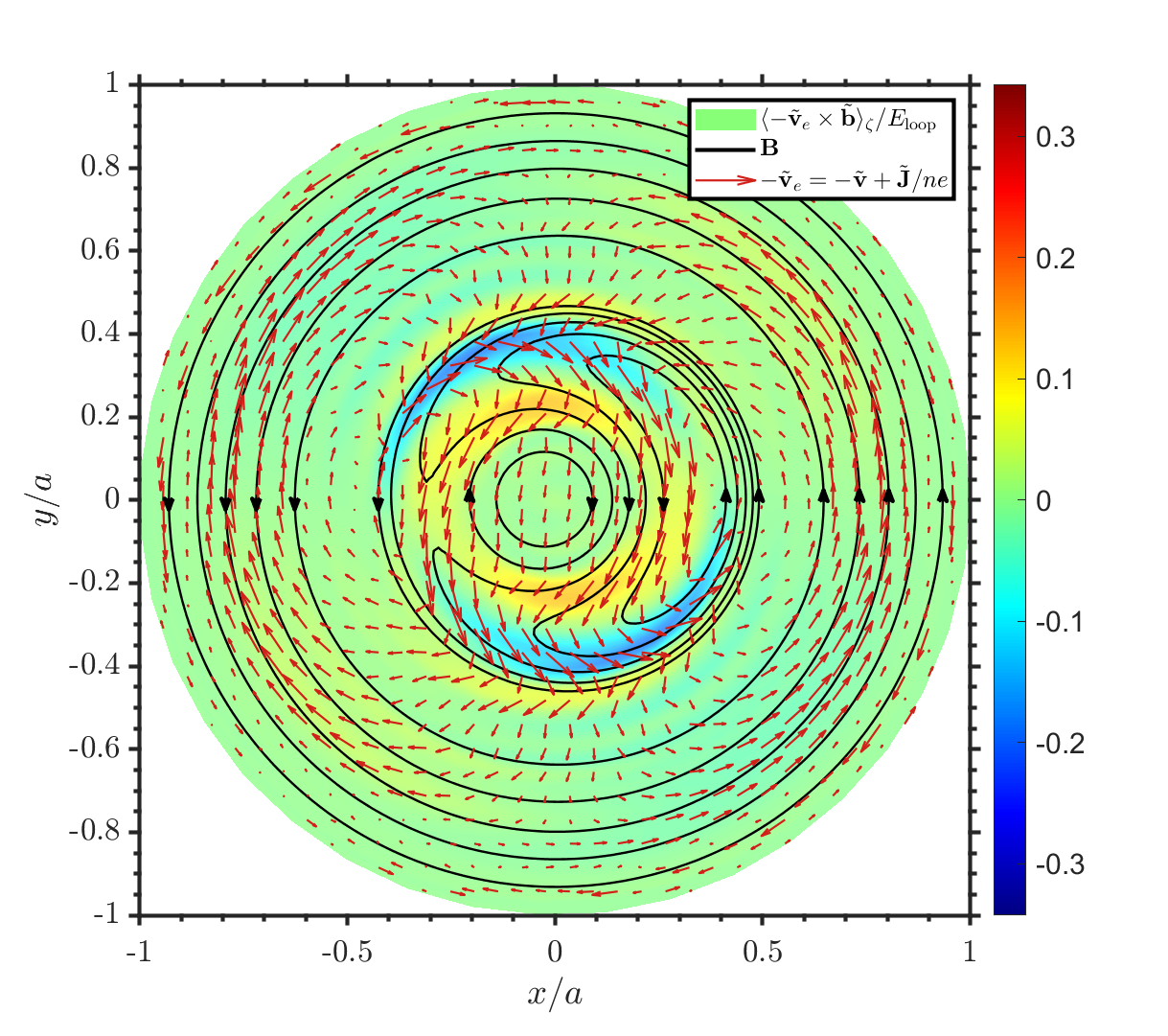}}
	\subfloat[]{\includegraphics[width=.45\textwidth]{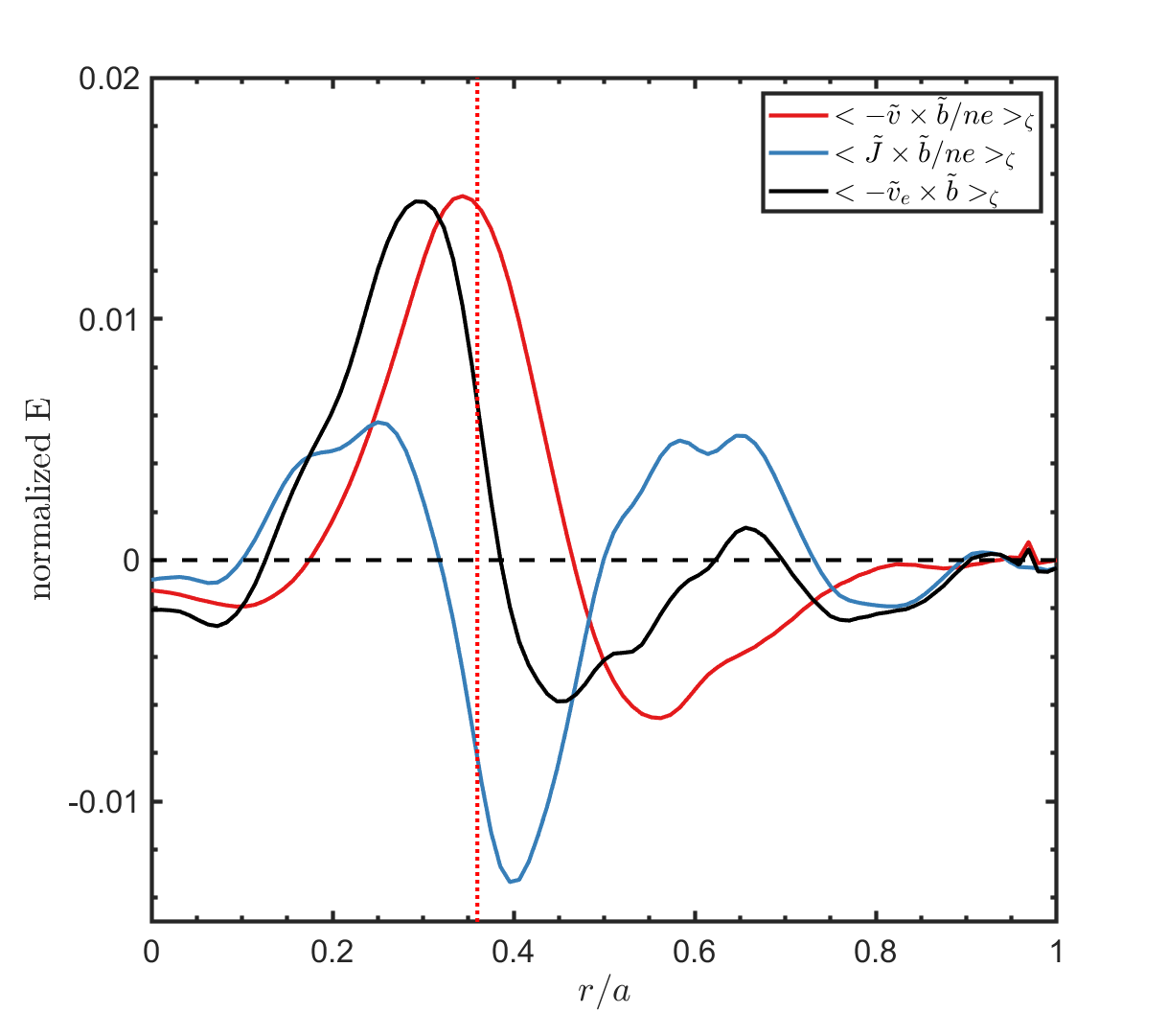}}
    \caption{Helical projection of the $(-1,7)$ mode in the two-fluid case. (a) Perturbed flow $\boldsymbol{\tilde{v}}_\zeta$ and the corresponding parallel MHD dynamo field. (b) Perpendicular current perturbation associated with the Hall electric field. (c) The electron flow $-\boldsymbol{v}_e = -\boldsymbol{v} + \boldsymbol{J}/ne$ and the total dynamo electric field. (d) Radial profiles of the mean MHD, Hall, and total dynamo fields. The equilibrium magnetic field is used only to outline the island structure; the dynamo electric field is computed from the perturbed magnetic field $\tilde{\boldsymbol{b}}$ alone.}  \label{fig:2flTMContour}
\end{figure}

For the two-fluid case shown in Fig. \ref{fig:2flTMContour} panel (a), the drift-tearing-like structure arises from the out-of-phase components in the presence of the Hall term and ion gyroviscosity.
The perturbed current  serves as the source of the Hall electric field. As shown in Fig. \ref{fig:2flTMContour}(b), the corresponding Hall dynamo field is strongly localized near the resonant surface and produces a pronounced negative contribution to the parallel electric field.
Fig. \ref{fig:2flTMContour}(c) shows the total dynamo electric field obtained by summing the MHD and Hall contributions, together with the electron flow $-\boldsymbol{v}_e = -\boldsymbol{v} + \boldsymbol{J}/ne$. Fig. \ref{fig:2flTMContour}(d) compares the radial profiles of the mean MHD, Hall, and total dynamo fields. The negative Hall peak near $r/a \simeq 0.37$ shifts the MHD peak inward; nevertheless, the total dynamo field remains positive inside and negative outside the resonant surface, consistent with the single-fluid result.

In summary, two-fluid effects introduce a phase shift between the perturbed flow and the magnetic island near the resonant surface. The Hall contribution is strongly localized, yet the overall tearing-mode dynamo profile remains qualitatively similar to that of the single-fluid case.

\section{Hall-Mediated Nonlinear Coupling and Breakdown of flux-pumping}

The preceding sections show that two-fluid effects modify the quasi-linear dynamo electric field and the helical structure of the saturated tearing modes, but these changes alone may not explain the transition to quasi-periodic relaxation. We therefore use a mode-resolved energy balance to separate quasi-linear drive, nonlinear coupling, and dissipation, and to compare the Hall and MHD transfers to secondary and higher-$n$ modes. This analysis shows how Hall-mediated energy redistribution disrupts the tearing-mode structure that maintains flux-pumping.

 The total energy of the $(m,n)$ component $\mathcal{U}_{mn}$ is defined as the sum of its kinetic and magnetic energies,
\begin{equation}
\mathcal{U}_{mn}=\mathcal{K}_{mn}+\mathcal{M}_{mn}
=\int_V\left(\frac{\rho |\boldsymbol{v}_{mn}|^2}{2}
+\frac{|\boldsymbol{b}_{mn}|^2}{2\mu_0}\right)\,\mathrm{d}V .
\label{eqn:ModalEnergy}
\end{equation}
Following the modal-energy diagnostic of Ho and Craddock \cite{Ho-1991}, the energy evolution of each Fourier component is divided into quasi-linear, nonlinear, and dissipative contributions. The corresponding modal power balance is written as
\begin{equation}
\Omega_{mn}\equiv\frac{\mathrm{d}\mathcal{U}_{mn}}{\mathrm{d}t}
=\Omega_{\mathrm{ql},mn}+\Omega_{\mathrm{nl},mn}
+\Omega_{\mathrm{diss},mn} .
\label{eqn:ModalPower}
\end{equation}
Here, $\Omega_{\mathrm{ql},mn}$ describes the energy exchange with the axisymmetric field, $\Omega_{\mathrm{nl},mn}$ represents the triadic transfer among non-axisymmetric modes, and $\Omega_{\mathrm{diss},mn}$ includes the resistive, viscous, and gyroviscous contributions. The weak electron-pressure and electron-inertia terms are neglected. The superscript \textit{``non''} denotes convolution over mode pairs satisfying the Fourier selection rules. The Lorentz, MHD dynamo, and Hall transfers are $\boldsymbol{v}_{mn}^{*}\cdot[\boldsymbol{j}\times\boldsymbol{b}]^{\mathrm{nonl}}_{mn}$, $\boldsymbol{j}_{mn}^{*}\cdot[\boldsymbol{v}\times\boldsymbol{b}]^{\mathrm{nonl}}_{mn}$, and $-\boldsymbol{j}_{mn}^{*}\cdot[\boldsymbol{j}\times\boldsymbol{b}]^{\mathrm{nonl}}_{mn}/(ne)$, respectively.
Normalizing by $\mathcal{U}_{mn}$ gives $\gamma_{\mathrm{tot},mn}=\Omega_{mn}/\mathcal{U}_{mn}$, $\gamma_{\mathrm{lin},mn}=(\Omega_{\mathrm{ql},mn}+\Omega_{\mathrm{diss},mn})/\mathcal{U}_{mn}$, and $\gamma_{\mathrm{nonl},mn}=\Omega_{\mathrm{nl},mn}/\mathcal{U}_{mn}$. A positive $\gamma_{\mathrm{nonl},mn}$ denotes net energy transfer to the target mode, while a negative value denotes transfer from it. Thus, the nonlinear growth rate measures energy redistribution rather than production; apart from boundary fluxes and numerical truncation, the ideal MHD and Hall transfers vanish when summed over all Fourier modes.

\begin{figure}[htbp]
	\centering
	\includegraphics[width=\textwidth]{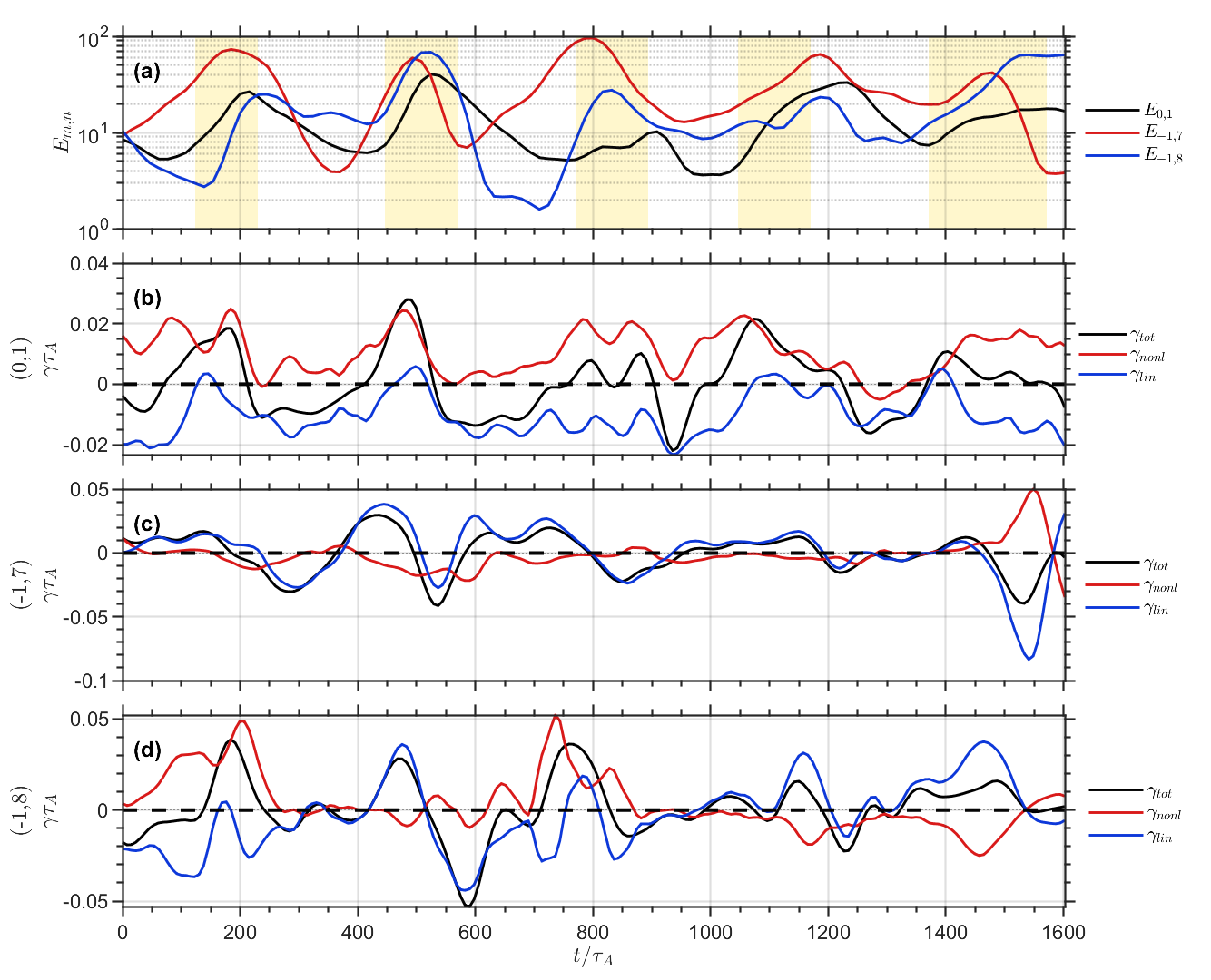}
	\caption{Modal-energy evolution and energy-growth-rate decomposition in the two-fluid simulation. (a) Total kinetic-plus-magnetic energies $\mathcal{U}_{mn}$ of the $(0,1)$, $(-1,7)$, and $(-1,8)$ components; the shaded intervals denote the phases of sawtooth events. (b)--(d) Total (black), nonlinear (red), and linear (blue) energy growth rates normalized by $\mathcal{U}_{mn}$. The $(0,1)$ component is driven predominantly by nonlinear transfer, while the nonlinear contributions to the primary tearing modes alternate between energy gain and loss.}
	\label{fig:ModesNonlinearGrowth}
\end{figure}

Fig.~\ref{fig:ModesNonlinearGrowth} compares the modal energies and energy growth rates of the $(0,1)$, $(-1,7)$, and $(-1,8)$ components. For the $(0,1)$ mode, $\gamma_{\mathrm{lin},mn}$ is negative over most of the interval, whereas $\gamma_{\mathrm{nonl},mn}$ is predominantly positive, showing that its recurrent growth is sustained by nonlinear transfer. The nonlinear contributions to the primary $m=-1$ tearing modes alternate in sign, indicating repeated intermode exchange. These bursts coincide with the sawtooth events, consistent with the enhanced three-wave coupling and spectral broadening observed experimentally \cite{Assadi-1992}. The $(0,1)$ mode is a typical nonlinear secondary mode whose growth accompanies perturbation localization and formation of a slinky-like structure through coupling of neighboring tearing modes \cite{Kusano-1987,Fitzpatrick-1999}. Together with the reduced eigenmode growth rate in Fig.~\ref{fig:LinearGrowth}, these results identify nonlinear transfer as the process governing departure from the flux-pumping state.

\begin{figure}[htbp]
	\centering
	\includegraphics[width=\textwidth]{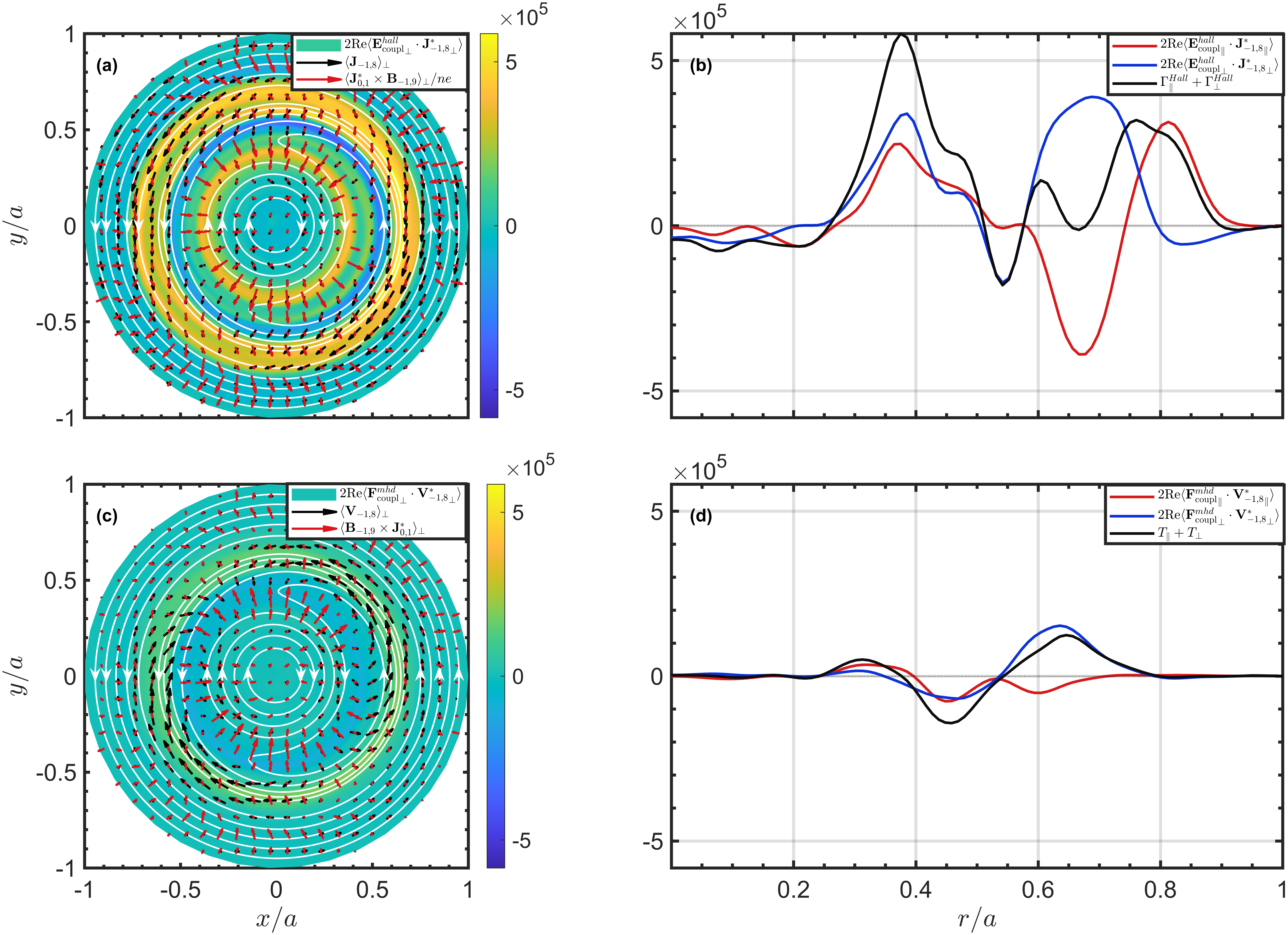}
	\caption{Spatial structure of a representative three-wave interaction during the nonlinear coupling phase centered at $t/\tau_A\simeq1400$, shortly before the fourth sawtooth crash. Panels (a) and (c) show helical projections of the Hall- and MHD-channel transfer densities, respectively, together with the interacting perturbation vectors and helical-flux contours. Panels (b) and (d) show the corresponding parallel, perpendicular, and total radial transfer profiles. For this triad and time interval, both the local amplitude and the net contribution of the Hall transfer exceed those of the MHD transfer.}
	\label{fig:ThreeWaveCoupling}
\end{figure}

The quantities plotted in Fig.~\ref{fig:ThreeWaveCoupling} are the local transfer densities associated with individual terms in $\Omega_{\mathrm{nl},mn}$. For the Hall interaction shown here, the plotted quantity may be written as
\[
\mathcal{I}_{\mathrm{H}}=\frac{1}{ne}
(\boldsymbol{j}_{0,1}\times\boldsymbol{b}_{-1,8})\cdot
\boldsymbol{j}_{-1,7}^{*}
=-\frac{1}{ne}
(\boldsymbol{j}_{-1,7}^{*}\times\boldsymbol{b}_{-1,8})\cdot
\boldsymbol{j}_{0,1}
\]
following the three-wave coupling rule $(1-1,-7+8)=(0,1)$. Including the sign of the Hall term, $-\operatorname{Re}\int_V\mathcal{I}_{\mathrm{H}}\,\mathrm{d}V$ gives the triad contribution to the nonlinear modal power and shows the exchange between the $(-1,7)$ and $(0,1)$ modes mediated by $\boldsymbol{b}_{-1,8}$. The MHD transfer follows by replacing $\boldsymbol{j}_{0,1}/(ne)$ with $\boldsymbol{v}_{0,1}$.

The selected interval is centered around $t/\tau_A\simeq1400$, shortly before the fourth sawtooth crash. For this triad, both the local amplitude and net Hall transfer exceed the corresponding MHD transfer. However, the transfers vary strongly in time, and different triads transfer energy both to and from the $(0,1)$ mode. After all modes and times are included, the net Hall and MHD contributions to the $(0,1)$ mode are comparable. The Hall interaction therefore produces strong bidirectional exchange, consistent with experimental evidence that three-wave coupling enhances the Hall dynamo by changing the phase relation between current and magnetic perturbations \cite{Ding-2006}.

\begin{figure}[htbp]
	\centering
	\includegraphics[width=0.6\textwidth]{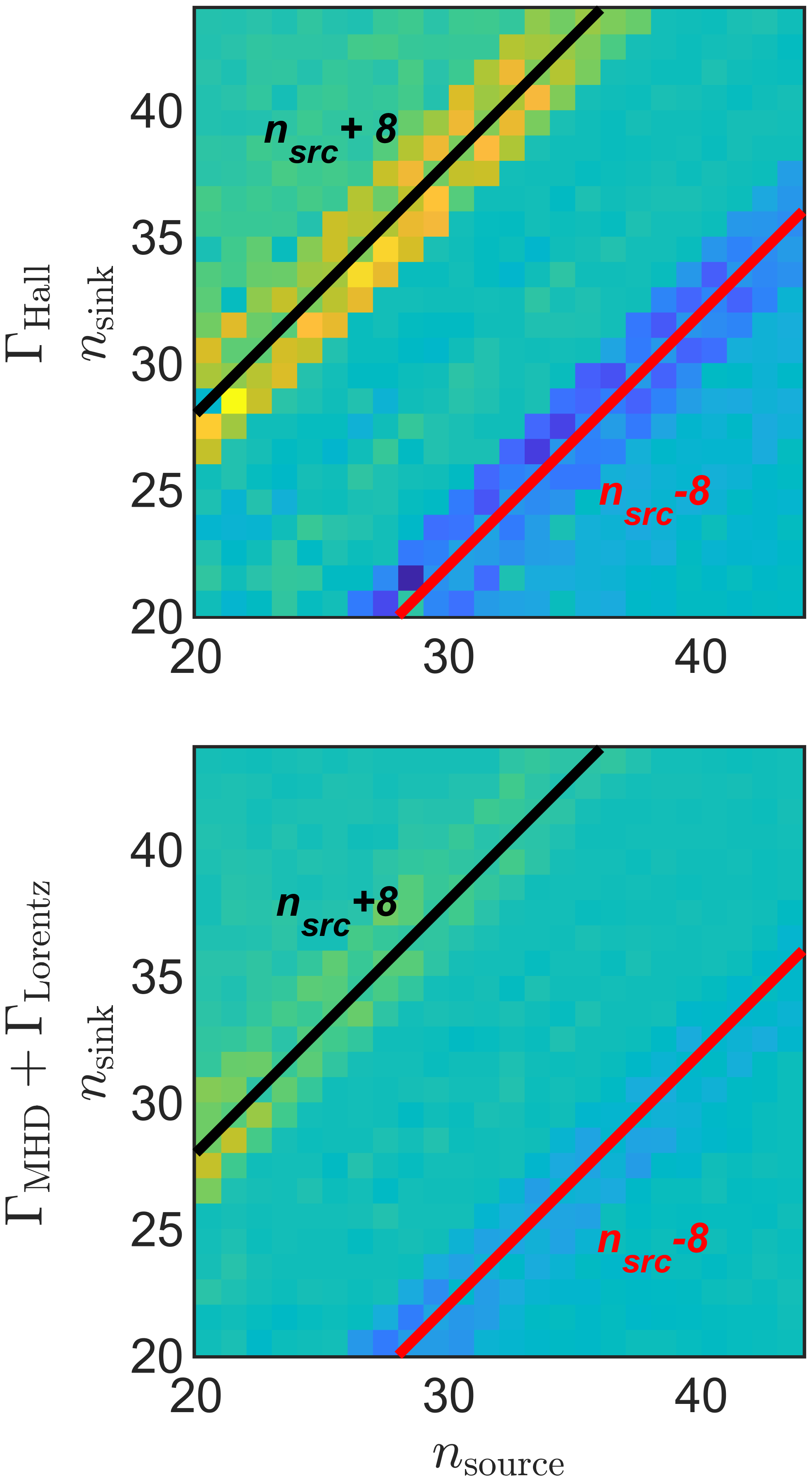}
	\caption{Nonlinear energy-transfer matrices in the high-$n$ band for (a) the Hall channel and (b) the combined MHD and Lorentz channels.  The paired interaction bands are consistent with a redistribution from lower-$n$ toward higher-$n$ components through coupling mediated by the dominant tearing modes. The larger Hall-channel amplitude identifies it as an important pathway for transfer toward shorter toroidal scales.}
	\label{fig:HigherBandTransfer}
\end{figure}

The transfer matrices in Fig.~\ref{fig:HigherBandTransfer} show the redistribution in mode-number space. In an RFP, the dominant $m=-1$ tearing modes have $n_t\simeq2R/a$ \cite{Marrelli-2021}; here, $2R/a\simeq7.4$, consistent with the dominant $n=7$--9 modes. The three-wave selection rule therefore gives coupling bands near $n_{\mathrm{out}}=n_{\mathrm{in}}\pm8$. These bands show transfer from lower-$n$ to higher-$n$ components through both the Hall and MHD/Lorentz channels, with a substantially larger Hall contribution. Similar $m=0$-mediated spectral broadening and crash-enhanced coupling have been reported in RFP simulations and MST experiments \cite{Holmes-1988,Assadi-1992}. The higher-$n$ components promote stronger localization and steeper current and magnetic-field gradients.

Taken together, the results distinguish two nonlinear responses. With weak intermode transfer, the saturated $m=-1$ tearing modes remain coherent and their quasi-linear dynamo maintains the steady flux-pumping \cite{Ho-1991}. Stronger two-fluid effects enhance the Hall-mediated exchange, drive the $(0,1)$ and higher-$n$ components, and disrupt this tearing-mode structure \cite{Fitzpatrick-1999}; the external current drive then rebuilds the current gradient and the cycle repeats. We refer  this transition as a Hall-mediated dynamical bifurcation between the steady flux-pumping and the quasi-periodic sawtooth relaxation. It represents two nonlinear states in the present scan of two-fluid effects, instead of  a change in the linear tearing instability. Hall-mediated spectral redistribution, rather than enhanced linear growth or a dominant net Hall drive of the $(0,1)$ mode, is therefore the primary process promoting sawtooth relaxation.

\section{Conclusion and Discussion}

This study investigates two-fluid effects on sawtooth relaxation in an RFP plasma in the low-Lundquist-number MH regime. The single-fluid simulations evolve toward a sawtooth-free state in which saturated $m=-1$ tearing modes generate a quasi-linear dynamo electric field that broadens the current profile and maintains flux-pumping. In contrast, the two-fluid simulations exhibit quasi-periodic sawtooth crashes. Radial profiles and helical projections show a strong local Hall dynamo near the resonant region, but spatial cancellation limits its contribution to the mean dynamo balance. The quasi-linear modification alone therefore cannot explain the transition between the two states.

A scan of the two-fluid effects shows increasing sawtooth activity despite decreasing linear tearing-mode growth rates. The modal-energy analysis instead reveals nonlinear growth of the $(0,1)$ mode and bidirectional exchange among the tearing modes. For the selected triad, both the local amplitude and net Hall transfer exceed the MHD transfer, but strong temporal variation and oppositely directed triads make the summed Hall and MHD contributions to the $(0,1)$ mode comparable. The Hall channel also transfers energy efficiently toward higher-$n$ modes. These results support a Hall-mediated dynamical bifurcation: weak nonlinear transfer maintains steady flux-pumping, whereas stronger Hall-mediated redistribution disrupts the coherent tearing-mode structure and produces quasi-periodic sawtooth relaxation. This bifurcation represents two nonlinear responses rather than a change in the underlying linear instability.

The present simulations also have important limitations. The Lundquist and Reynolds numbers are of order $10^4$, corresponding to a substantially more dissipative regime than that of present RFP experiments \cite{Ding-2004,Liu-2019}. Under these conditions, the ability of the MH tearing modes to maintain a sawtooth-free flux-pumping state may differ significantly from that under experimental conditions. At higher $S$, strong nonlinear coupling and sawtooth activity may already occur in the single-fluid model, without the same transition observed in the present low-$S$ scan. Whether two-fluid effects remain comparably important for intermode energy transfer in experimentally relevant resistivity regimes therefore remains an open question.

Future work will extend the parameter scan toward experimentally relevant Lundquist numbers and include a more realistic ion-temperature evolution. Mode-resolved energy-transfer diagnostics in both simulations and experiments will be required to determine how the Hall term and ion gyroviscosity modify nonlinear coupling, spectral redistribution, and magnetic reconnection in the high-$S$ regime of RFP experiments.



\section*{Acknowledgment}

The authors express sincere thanks to all members of the KTX group as well as the NIMROD team. This work was supported by the National Natural Science Foundation of China (Grant No. 11775220), the National Magnetic Confinement Fusion Energy Development Program of China (Grant Nos. 2017YFE0301702 and 2019YFE03050004), and U.S. Department of Energy (Grant No. DE-FG02-86ER53218). The computing work in this paper was supported by  the Supercomputing Center of USTC and the Public Service Platform of High Performance Computing by Network and Computing Center of HUST.

\newpage

\section*{References}

\bibliographystyle{unsrt}
\bibliography{References} 

\newpage

\newpage

\newpage

\newpage



\newpage



\newpage

\newpage

\newpage

\newpage

\newpage

\newpage

\newpage

\newpage
\newpage

\newpage

\newpage

\newpage

\end{document}